\documentclass[pra,floatfix,showpacs,amsmath]{revtex4}

\makeatletter
\let\@ORGREVTEXendnotemark\@endnotemark
\let\@ORGREVTEX@makefnmark@cite\@makefnmark@cite
\def\@endnotemark{\bgroup\@fileswfalse\@ORGREVTEXendnotemark\egroup}
\def\@makefnmark@cite{\bgroup\@fileswfalse\@ORGREVTEX@makefnmark@cite\egroup}
\makeatother

\usepackage{amssymb}
\usepackage[mathscr]{eucal}
\usepackage{graphicx}

\usepackage{braket}
\usepackage{enumerate}
\usepackage{subfigure}

\usepackage{mciteplus}
\mciteErrorOnUnknownfalse

\def\tr{\mbox{Tr}}

\def\vecmap{\mbox{vec}}

\newcommand{\bvec}[1]{\mathbf{#1}}

\def\a{{\bvec{a}}}
\def\A{{\bvec{A}}}

\def\x{{\bvec{x}}}
\def\y{{\bvec{y}}}

\renewcommand\Re{\operatorname{Re}}
\renewcommand\Im{\operatorname{Im}}

\begin{document}

\title{Exploiting translational invariance in Matrix Product State simulations
of spin chains with periodic boundary conditions}

\author{B. Pirvu$^1$, F. Verstraete$^1$, G. Vidal$^2$}
\affiliation{$^1$Fakult\"at f\"ur Physik, Universit\"at Wien,
Boltzmanngasse 5, A-1090 Wien, Austria\\
$^2$School of Mathematics and Physics, The University of Queensland,
QLD 4072, Australia}

\date{\today}

\begin{abstract}
We present a matrix product state (MPS) algorithm to approximate ground
states of translationally invariant systems with periodic boundary
conditions. For a fixed value of the bond dimension $D$ of the MPS, we
discuss how to minimize the computational cost to obtain a seemingly
optimal MPS approximation to the ground state. In a chain of $N$ sites
and correlation length $\xi$, the computational cost formally scales as
$g(D,\xi/N)D^3$, where $g(D,\xi/N)$ is a nontrivial function. For
$\xi \ll N$, this scaling reduces to $D^3$, independent of the system
size $N$, making our algorithm $N$ times faster than previous proposals.
We apply the method to obtain MPS approximations for the ground states of
the critical quantum Ising and Heisenberg spin-$1/2$ models as well as
for the noncritical Heisenberg spin-$1$ model. In the critical case, for any
chain length $N$,
we find a model-dependent bond dimension $D(N)$ above which the polynomial
decay of correlations is faithfully reproduced throughout the entire system.
\end{abstract}

\maketitle



\section{Introduction}

Concepts of entanglement for many-body quantum systems have recently
proven useful to devise new methods for the numerical simulation of
quantum spin chains. It has been shown that the very successful density
matrix renormalization method (DMRG)~\cite{white-1992} can be rephrased
as a variational method over the class of matrix product states
(MPS)~\cite{rommer-ostlund-1997,vidal-2004,frank-2004a,frank-2008};
this realization clarified
the relatively poor performance of DMRG for systems with periodic
boundary conditions (PBC), as MPS with open boundary conditions (OBC)
do not have the right entanglement structure.
It was shown in~\cite{frank-2004a} how this
could be cured by using a MPS with PBC. However, due to the cyclic
structure of the underlying MPS, the computational cost of the simulation
in terms of the MPS bond dimension $D$ grew from $O(D^3)$ to $O(D^5)$.
This was subsequently lowered to $O(D^3)$
in~\cite{sandvik-vidal-2007,pippan-2010}.

An important motivation to study finite chains is that one can compute
bulk properties of the system in the thermodynamic limit by extrapolating
results obtained for increasingly large chains \cite{book-cardy-1996}.
In this context, it is relevant whether OBC or PBC are considered. For a
finite chain with OBC, local expectation values differ from those in
thermodynamic limit due both to finite-size effects and to boundary effects,
and larger chains need to be considered. In contrast, with PBC only
finite-size effects are present. This makes the extrapolation to the
thermodynamic more transparent and smaller systems need to be simulated.
Another important advantage of PBC is that only in this case a finite chain
can be translation invariant (TI)
\nocite{nishino-okunishi-1995,*hieida-okunishi-akutsu-1997,
*okunishi-hieida-akutsu-1999,*ueda-nishino-okunishi-hieida-derian-gendiar-2006}
\nocite{vidal-2007,mcculloch-2008,orus-2008,me-2010}
\footnote{TI can also be exploited for MPS simulations with OBC, but this
requires addressing an infinite system
\cite{white-1992,nishino-okunishi-1995,
vidal-2007,mcculloch-2008,orus-2008,me-2010}.
Notice that since the system size $N$ is infinite from the start, there can
not be finite-size or boundary corrections to the bulk properties of the
system. However, in this case numerical results are contaminated by effects
due to the finite bound dimension $D$ of the MPS. Interestingly, one can
apply "finite-D" scaling techniques to extract accurate estimates of bulk
properties~\cite{nishino-1996,andersson-boman-ostlund-1999,tagliacozzo-2008,
moore-2008}.}.
This is crucial feature for the present work, where TI
\footnote{We will use the abbreviation TI to denote both the adjective
"translational invariant" and the noun "translational invariance".}
is exploited in order
to reduce the computational costs of simulating finite chains.

\begin{figure}[htbp]
  \centering
  \subfigure[~Medium $\xi$, small $l$: equivalent to OBC environment.]{
    \label{fig:Xi_effect_on_PBC_chains_a}
    \includegraphics[width=0.30\textwidth]
    {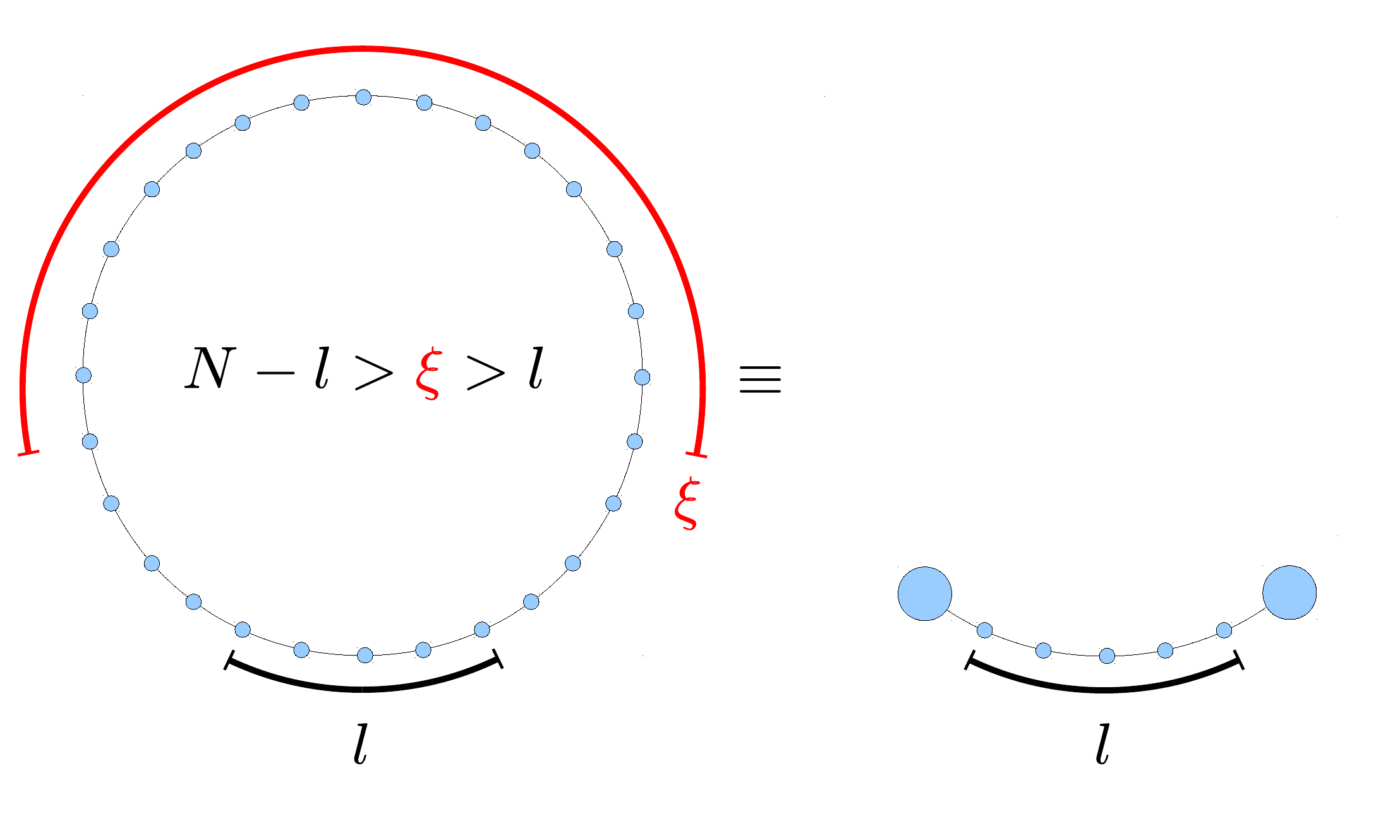}
  }
  \subfigure[~Medium $\xi$, medium $l$: equivalent to partially
             correlated OBC environment.]{
    \label{fig:Xi_effect_on_PBC_chains_b}
    \includegraphics[width=0.32\textwidth]
    {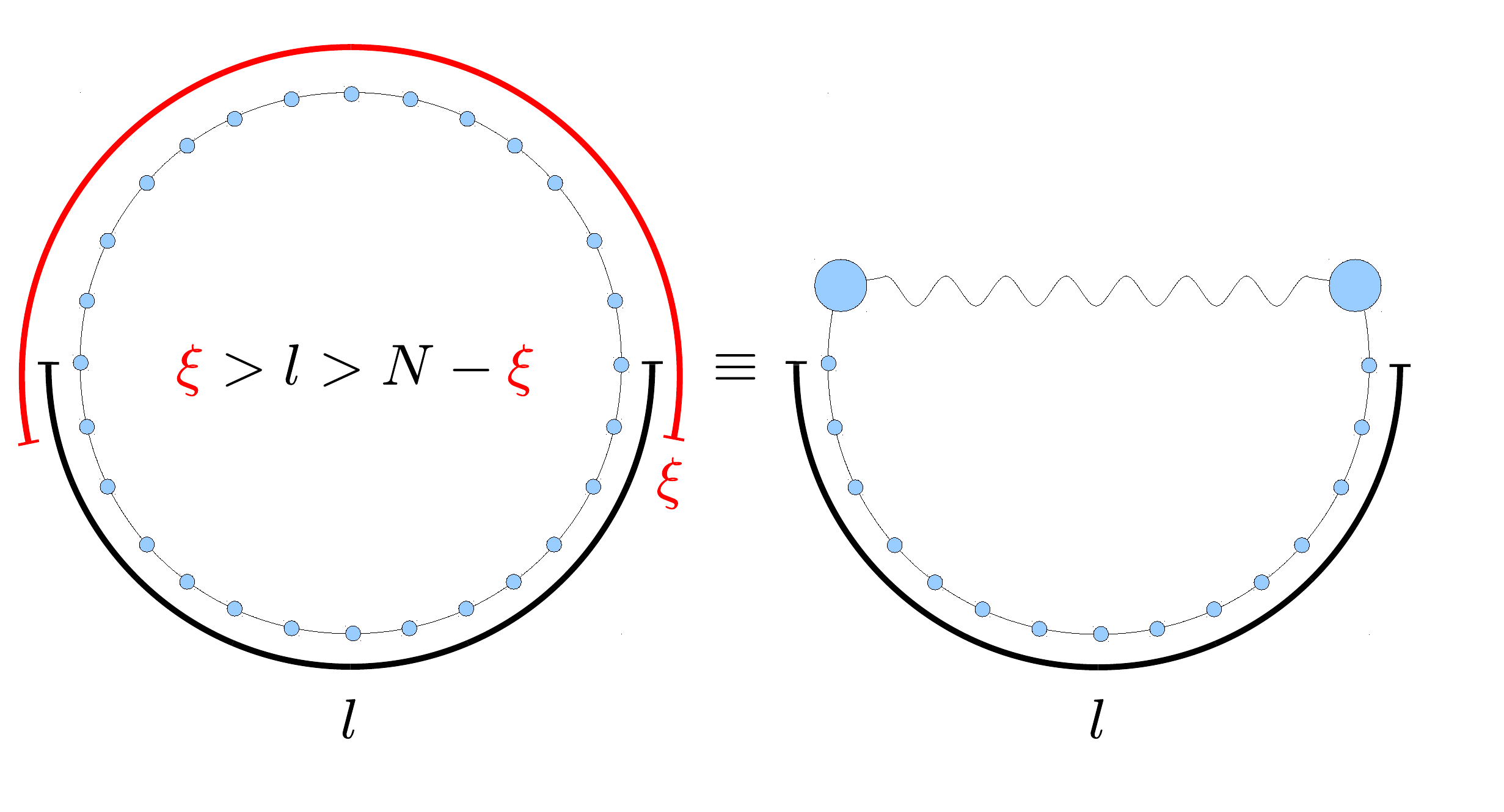}
  }
  \subfigure[~Large $\xi$, any $l$: equivalent to fully
             correlated OBC environment.]{
    \label{fig:Xi_effect_on_PBC_chains_c}
    \includegraphics[width=0.32\textwidth]
    {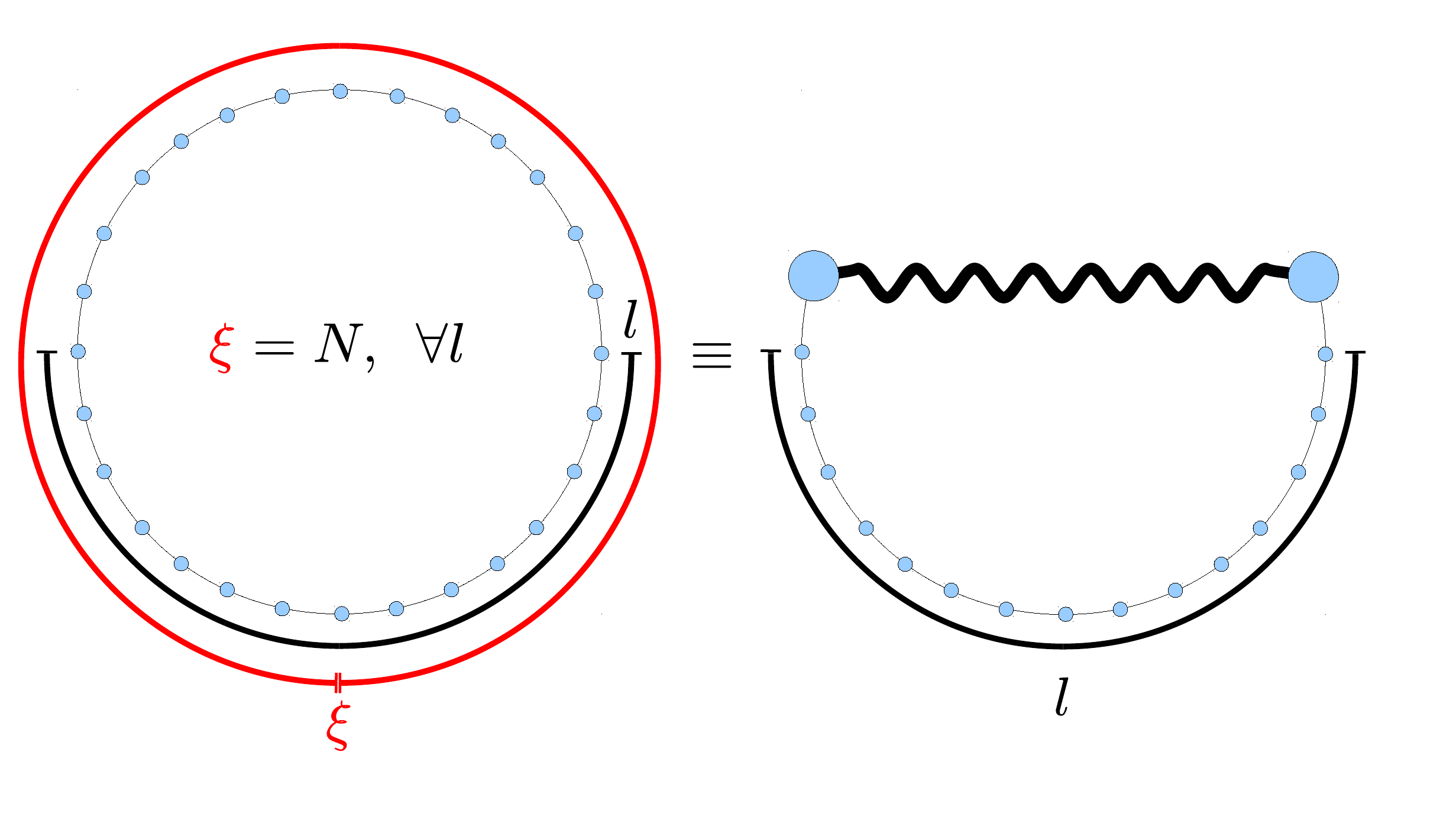}
  }
  \caption{(Color online).
           The properties of a block of size $l$ within a PBC system can be
           equivalent to those of a block with same size in the bulk of an
           OBC system. Depending on $l$ and the correlation length $\xi$,
           the left and right boundary conditions of the OBC system are
           more or less correlated.}
  \label{fig:Xi_effect_on_PBC_chains}
\end{figure}

Pippan, White and Evertz~\cite{pippan-2010} recently showed how to simulate
spin chains with PBC with an MPS algorithm whose computational cost given
in terms of $D$ scales like $O(D^3)$.
The intuition behind this scaling can be understood if one first considers
systems with a correlation length $\xi$ that is much shorter than the system
size $N$. Let us choose a block of sites with size $l$ such that $\xi>l$
(see figure ~\ref{fig:Xi_effect_on_PBC_chains}a). In this case correlations
between the left and the right ends of the block are mediated only through
the sites inside the block. It is clear that the properties of this block are
exactly the same as those of a block of equal length embeded in the bulk of a
sufficiently large system with OBC. It is then not surprising that computing
observables that are contained within the block has a cost proportional to
$D^3$, as in the case of OBC. This is basically due to the fact that such
calculations involve contracting a tensor network that has, as
\emph{uncorrelated} left and right boundary conditions, two
\emph{boundary vectors} with $D^2$ components \cite{frank-2008}.
Now imagine we are interested in the description of properties contained
in a larger block such that $\xi > l > N-\xi$ (see
figure~\ref{fig:Xi_effect_on_PBC_chains}b).
This block is small enough for its ends
to have correlations that are mediated via its own sites,
yet large enough that correlations are also mediated via the sites
outside the block, since now $N-l<\xi$. If these externally mediated
correlations are relatively small, the situation is not very different
from the previously described case where $l<N-\xi$. All we have to do
is to replace the two uncorrelated \emph{boundary vectors} with a low rank
\emph{boundary matrix} that contains the small amount of correlations.
If the rank of the matrix is $n$, then the cost of this algorithm will be
proportional to $nD^3$.

We emphasize two important aspects of the computational cost of the algorithm
in Ref.~\cite{pippan-2010}. The \emph{first} one is that the cost is also
proportional to the system size $N$, due to the usual sweeping procedure
that optimizes one site at each instant.
We will show below how, in the case of a TI chain, one can get rid of this
factor
\footnote{
For chains where $\xi \ll N$ the cost will not depend on $N$ at all.
If $\xi \approx N$ on the other hand the cost will contain a factor
that is smaller $N$ but is nevertheless an artifact thereof.
}.
This is achieved by
using a TI MPS, where the $N$ tensors of the MPS are chosen to be identical.
For all $D$, the precision of our results is comparable to that reported
in Ref.~\cite{pippan-2010}. This indicates that restricting the MPS ansatz
to be TI does not lead to a loss of precision, while yielding a substantial
reduction of the computational cost. The \emph{second} one is the
multiplicative factor $n$ corresponding to the rank of the boundary matrix
that transfers correlations between the ends of a block.
In the case where the correlation length $\xi$ is of the order of the system
size $N$ (see figure~\ref{fig:Xi_effect_on_PBC_chains}c), this factor may not
be small. In a worst case scenario, where strong correlations between distant
sites would force the boundary matrix to be full rank, i.e. $n=D^2$, the
approach in Ref.~\cite{pippan-2010} would not be better than the $O(D^5)$
algorithm of Ref.~\cite{frank-2004a}.
Thus for critical systems where $\xi \approx N$ it is a priori unclear what
the overall scaling of the computational cost in $D$ will be. 
However, in Ref.~\cite{pippan-2010} it has been indicated
that if $D$ is not too large, the ground state energy of a critical spin
chain obtained using a small constant $n$ is satisfactory, in that its
accuracy scales with $D$ in a similar way as it would in an OBC chain of
the same size.

Here we shall show how to exploit TI to obtain a faster algorithm that, for
instance, does not scale with $N$ when $\xi \ll N$. However, except for the
case $\xi\ll N$, we still lack a precise characterization of how the cost
scales as a function of $D$ and $N$. We benchmark the present approach by
addressing both critical (i.e. $\xi \approx N$) and non-critical
(i.e. $\xi\ll N$) chains. An important observation is that in the case
of critical systems the finite bond dimension $D$ of the MPS introduces
an effective correlation length $\tilde{\xi}_D \approx D^{\kappa}$
\cite{nishino-1996,andersson-boman-ostlund-1999,tagliacozzo-2008,moore-2008}
that depending on $D$ can be much smaller than the actual one.
This implies that as $N$ grows, a larger bond dimension
$D \approx N^{1/\kappa}$ needs to be considered if correlations between
distant sites of the chain with PBC are to be properly captured
Our numerical results are consistent with a complex scenario where the cost
of simulations is dominated by the crossover between finite-$N$ and
finite-$D$ corrections, as further discussed in~\cite{inprep-2010}.

The rest of the paper is structured as follows: we start by sketching the
main idea of the approach in Sect. \ref{sec:overview}, followed by an
in-depth presentation of the algorithm in Sect. \ref{sec:algorithm}. In
Sect. \ref{sec:numerical_results} we present numerical results for the
critical Quantum Ising and Heisenberg spin-$1/2$ models as well as
for the non-critical Heisenberg spin-$1$ model.
Finally Sect. \ref{sec:conclusions} contains some conclusions.

\section{Overview}
\label{sec:overview}

This work is concerned with the approximation of ground states (GS) within
the variational class of MPS with PBC defined in~\cite{frank-2004a}.
Since critical systems are arguably among the most challenging ones from a
computational perspective, we will apply the approach to investigate
critical spin chains (although non-critical chains can also be considered).
An important restriction is that we only consider TI systems, which we will
analyse with a TI MPS ansatz, namely an MPS where the tensors corresponding
to different sites are all equal. The resulting variational class is a
subclass of the one defined in \cite{frank-2004a}. The TI MPS with PBC reads
\begin{equation}
  \ket{\psi(A_i)}=
  \sum_{i_1,\dots i_N=1}^{d}
  \tr\big(A_{i_1} A_{i_2} \dots A_{i_N}\big)
  \ket{i_1 i_2 \dots i_N}
\end{equation}
\noindent
with identical matrices $A_i$ at every site. Note that since for fixed $i$,
each $A_i$ represents a matrix, the MPS is completely characterized by the
three dimensional tensor
$A_{i\phantom{\alpha}\beta}^{\phantom{i}\alpha}=:\bvec{A}$.
Furthermore we should point out that we will mostly be interested in
Hamiltonians that are real and reflection invariant; these symmetries can be
implemented at the level of MPS by choosing the matrices $A_i$ real and
symmetric. This extra constraint does not seem to deteriorate the accuracy
of the variational procedure.

Since our ansatz consists of $N$ copies of the same tensor, the energy is not
a quadratic expression in the variables defined by the tensors $A_i$; this
implies that we cannot use the sweeping procedure described
in~\cite{frank-2004a} or any other procedure that
lowers the energy by minimizing it for one site at a time. While this might
seem a reason to be concerned at first, it will actually be the key to
reducing computational costs.

The advantages of a TI MPS ansatz (with periodicity one or two) have already
been exploited in the context of infinitely long chains~\cite{white-1992,
nishino-okunishi-1995,vidal-2007,mcculloch-2008,orus-2008,me-2010}.
Refs.~\cite{white-1992,nishino-okunishi-1995,mcculloch-2008}
used a TI MPS in the context of infinite system DMRG.
In Ref.~\cite{vidal-2007}, instead, a
(two-site periodic) MPS approximation to ground states was obtained by
imaginary time evolution. Refs.~\cite{orus-2008, me-2010} discussed how
to compute ground states with a one-site TI MPS when the imaginary time
evolution operator can be well enough approximated by layers of
one-site TI matrix product operators. An attempt to adapt that method to
finite chains with PBC yielded results that are not as accurate as one might
expect
\footnote{This is basically due to the fact that the bond dimension
truncation method used in \cite{orus-2008, me-2010} can be shown to be
optimal (in a certain sense) only for infinitely long chains.
We have used a straightforward adaptation of that method for finite
chains with PBC and the results are between one and a few orders of
magnitude worse than the ones obtained by the gradient method described
in this work.}.
Finally, we also point out that a TI MPS with PBC was
already used in Ref.~\cite{sandvik-vidal-2007} together with Monte Carlo
sampling techniques, with a formal cost $O(ND^3)$. In that case, the use
of sampling techniques reduced the cost from $O(D^5)$ to $O(D^3)$, but at
the same time enforced the multiplicative factor $N$, since a TI MPS does
not represent a TI state once a given configuration is chosen during the
sampling.

An obvious way to find the TI-MPS with minimal energy is a multidimensional
minimization procedure that requires
only evaluations of the function itself, such as the downhill simplex
method~\cite{numrec-2007}.
When no further information about the function is available, this is indeed
the method of choice. It is extremely robust but also extremely slow.
However, if there is a feasible way to obtain more elaborate information such
as the gradient or the Hessian, there are methods relying on these quantities
that are clearly superior in what regards the speed of convergence and the
required storage space.

In the following we will present an efficient algorithm to calculate
the gradient of the energy $\nabla E(\bvec{a})$
where the argument $\bvec{a}=\vecmap(\bvec{A})$
denotes the vector containing all entries of the MPS tensor $\bvec{A}$.
The result will then be used by a standard numerical library conjugate gradient
algorithm to find a minimum of $E(\bvec{a})$. We must emphasize that this
minimum is by no means guaranteed to be the global one i.e. the optimal
ground state approximation within the subspace defined by our special
MPS ansatz. However, our numerical results seem to be slightly more accurate
than previous results \cite{pippan-2010}, while we have obtained a reduction
in computational costs.
We will illustrate the accuracy of this approach by applying it to two
exactly solvable models in order to give exact values for the numerical errors.

The computational cost will turn out to scale as $O(mnD^3)+O(n^2D^3)$
where $D$ is the virtual bond dimension and $m$ and $n$ are some parameters
to be specified below.
Briefly speaking, the scaling can be understood as follows:
first we approximate large powers of the MPS transfer matrix,
whose exact definition will be given later in the text,
within a reduced subspace of dimension $n$.
Treating each of the $n$ dimensions separately allows us to
transform the contraction of a tensor network with PBC
(which scales as $O(D^5)$) into $n$ contractions
of tensor networks with OBC (each of which scales as $O(D^3)$).
As we will explain in more detail in the next section, the resulting
tensor networks will still contain at most one portion represented
by say $m$ adjacent transfer matrices that is not connected to
the already approximated one. If $m$ is large, this second
portion can again be approximated within a $n$-dimensional subspace
thereby yielding the scaling $O(n^2D^3)$.
If $m$ is small, we are forced to contract the transfer matrices one
after the other which gives the scaling $O(mnD^3)$.

\section{The algorithm}
\label{sec:algorithm}

Let us rearrange the MPS tensor components in a vector
$\bvec{a}=\vecmap(\bvec{A})$
which allows us to write the energy as a function over the manifold of
free parameters in the MPS

\begin{equation}\label{eq:Eexpval_vecA}
  E(\bvec{a})
        =\frac{\bra{\psi(\bvec{a})} H \ket{\psi(\bvec{a})}}
              {\braket{\psi(\bvec{a})|\psi(\bvec{a})}}
        \equiv\frac{\bra{\psi(\bvec{A})} H \ket{\psi(\bvec{A})}}
              {\braket{\psi(\bvec{A})|\psi(\bvec{A})}}
        \,\,\,.
\end{equation}
\noindent
Note that due to the constraints that the matrices are real and symmetric,
the number of vector components in $\bvec{a}$ has been reduced to
$\frac{1}{2}dD(D+1)$. As we will treat only spin-1/2 chains
(i.e. $d=2$) in this work, the variational parameter manifold is actually
$D(D+1)$-dimensional.
Furthermore we will denote expectation values taken with respect
to the MPS defined by the tensor $\bvec{A}$ as
$\langle O\rangle_\bvec{A}:=\bra{\psi(\bvec{A})} O \ket{\psi(\bvec{A})}$.

Also note that (\ref{eq:Eexpval_vecA}) can have local extrema as opposed to
$E(\Psi)=\bra{\Psi}H\ket{\Psi}$ which is a convex quantity in the
exponentially large Hilbert space.
The MPS-parametrization restricts
the full parameter space to a submanifold thus possibly generating
local extrema where all derivatives in this subspace vanish.
If one uses as a starting point of the conjugate gradient
algorithm a random vector $\bvec{a}_{rand}$, the search algorithm will
typically get stuck in a local minimum.
In order to avoid getting stuck in one of these, we will choose as a
starting point a vector $\bvec{a}_0$ of which we can be sure that it
is close to the global minimum. This approach turns out to be very
robust and fast.
If we are interested in ground states of chains with very large $N$,
the most natural choice for the starting vector is
an MPS approximation of the GS of the same model in the thermodynamic
limit. Note that this MPS must have exactly the same symmetry
properties as our ansatz.
It was shown in previous work \cite{me-2010} how to obtain this MPS and
we will actually use the tensors computed there as starting points for
the present algorithm.
It is obvious why the MPS for the GS of the infinite chain is a
good choice if one is interested in finite PBC-chains with
$N\gg\xi(D)$, where $\xi(D)$ is the correlation length induced by finite
$D$. However, it turns out that this approach also works satisfactory
for moderately large $N$.
Of course, if there already is any PBC solution available, using that one
as a starting point may provide a gain in convergence time,
especially if the chain lengths are similar.

\begin{figure}[ht]
  \begin{center}
    \includegraphics[width=0.9\textwidth]
    {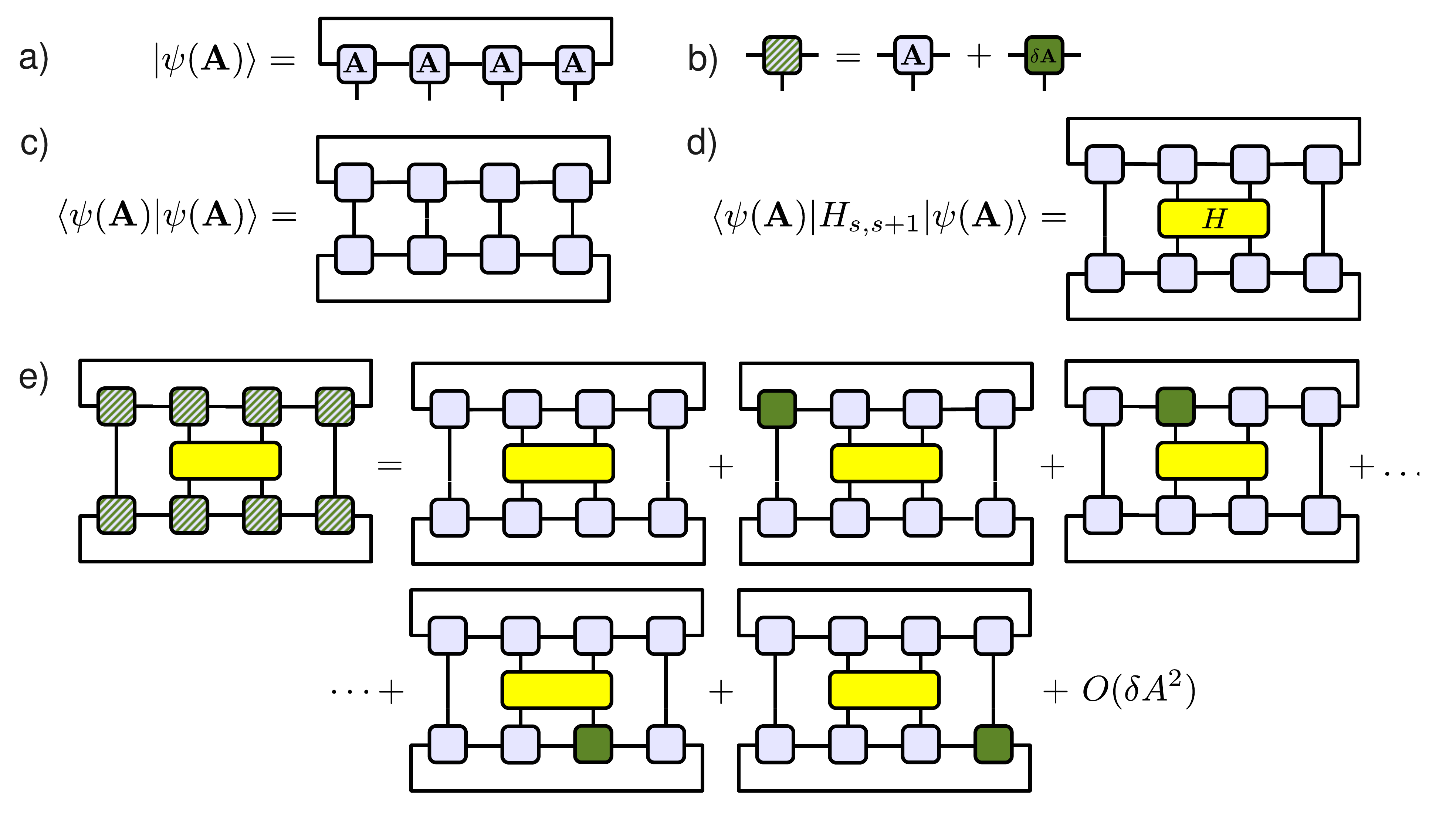}
  \end{center}
  \caption{
    (Color online). (a) Graphical representation of the TI PBC MPS
    $\ket{\psi(\bvec{A})}$ of a TI spin chain with 4 sites.
    Note the identical tensors $\A$ at each site.
    (b) Small perturbation $\delta\A$ is added to
    the to the MPS tensor $\A$.
    (c) Norm of a state $\braket{\psi(\A)|\psi(\A)}$.
    (d) Expectation value of a 2-site operator e.g.
    $\braket{\psi(\A)|H_{s,s+1}|\psi(\A)}$.
    (e) The expectation value is expanded in powers of $\delta\A$.
  }
\label{fig:PBC_expval_A_expansion_LARGE}
\end{figure}

The gradient $\nabla E(\bvec{a})$ reads explicitly

\begin{equation}\label{eq:grad_Eexpval_vecA}
  \nabla E(\bvec{a})=
  \frac{1}{\braket{\psi(\bvec{a})|\psi(\bvec{a})}}
  \nabla
  \bra{\psi(\bvec{a})} H \ket{\psi(\bvec{a})}
 -\frac{\bra{\psi(\bvec{a})} H \ket{\psi(\bvec{a})}}
        {\braket{\psi(\bvec{a})|\psi(\bvec{a})}^2}
  \nabla
  \braket{\psi(\bvec{a})|\psi(\bvec{a})} \,\,\,.
\end{equation}

It turns out that this quantity can be computed efficiently. First, since
we assume a translationally invariant Hamiltonian with nearest neighbour
interactions, we have
\[\langle H\rangle_\bvec{A}=\langle H_{N,1}\rangle_\bvec{A}+\sum_{s=1}^{N-1}
\langle H_{s,s+1}\rangle_\bvec{A} = N \langle H_{s,s+1}\rangle_\bvec{A}.\]
Hence the first term in (\ref{eq:grad_Eexpval_vecA}) is proportional to
the gradient of the energy density
$\rho_E(\bvec{a})=\langle H_{s,s+1}\rangle_\bvec{A}$, $\forall s\in[1,N]$
(see figure~\ref{fig:PBC_expval_A_expansion_LARGE}d).
Second, we can obtain gradients such as the ones occurring in
(\ref{eq:grad_Eexpval_vecA}) numerically at a given point
$\bvec{a}$ by expanding the differentiated quantity in powers of
$\delta \bvec{a}$ and computing the coefficient of the linear term.
Thus the derivative in the first term is obtained via

\begin{equation}\label{eq:Eexpval_expansion}
  \rho_E(\bvec{a}+\delta \bvec{a}) = \rho_E(\bvec{a}) +
    \delta \bvec{a} \bigl[\nabla_{\bvec{a}'}
    \rho_E(\bvec{a}')\bigr]_{\bvec{a}'=\bvec{a}} +
    O(\delta \bvec{a}^2)
\end{equation}
\noindent
and the one in the second via

\begin{equation}\label{eq:norm_expansion}
  \braket{\psi(\bvec{a}+\delta\bvec{a})|\psi(\bvec{a}+\delta\bvec{a})}
  = \braket{\psi(\bvec{a})|\psi(\bvec{a})}
  + \delta \bvec{a} \bigl[\nabla_{\bvec{a}'}
    \braket{\psi(\bvec{a}')|\psi(\bvec{a}')}
    \bigr]_{\bvec{a}'=\bvec{a}} +
    O(\delta \bvec{a}^2)\,\,\,.
\end{equation}
\noindent

\begin{figure}[ht]
  \begin{center}
    \includegraphics[width=0.9\textwidth]{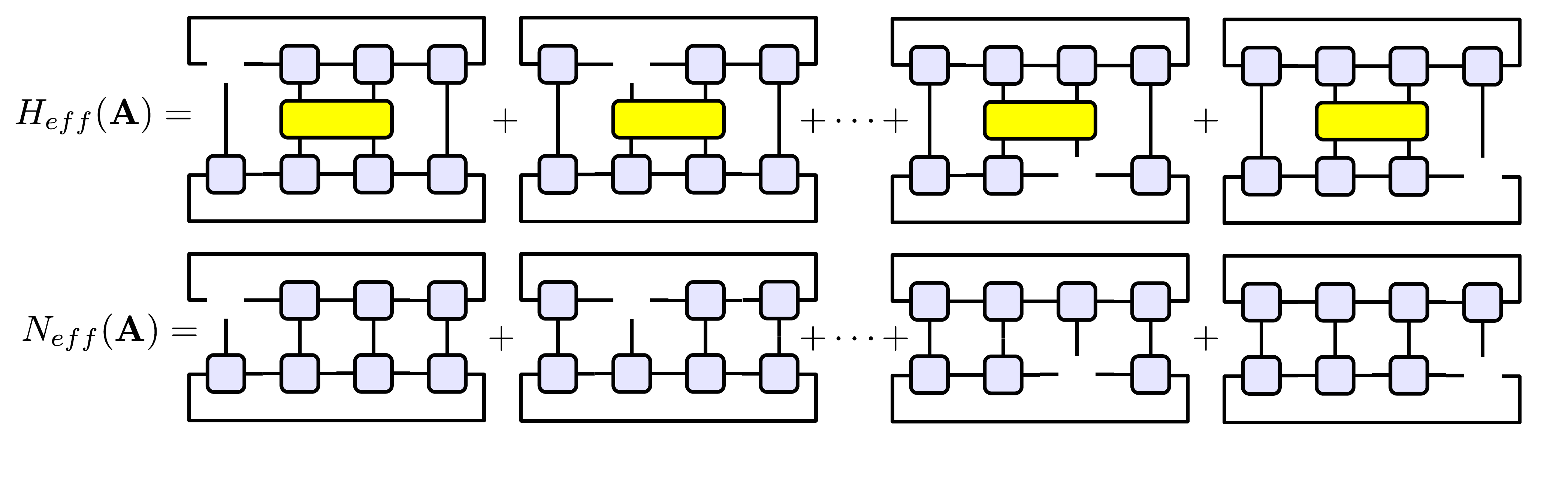}
  \end{center}
  \caption{
    (Color online). Graphical representation of the tensor
    $H_{eff}(\bvec{A})$ and $N_{eff}(\bvec{A})$.
  }
\label{fig:Heff_Neff}
\end{figure}

Let us first consider (\ref{eq:Eexpval_expansion}).
This can can be computed explicitly by taking
a sum of completely contracted tensor networks
(see figure~\ref{fig:PBC_expval_A_expansion_LARGE}e).
Let $H_{eff}(\bvec{A})$ denote the object that is obtained by removing the
tensor $\delta\bvec{A}$ from each term of $\rho_E(\bvec{a}+\delta \bvec{a})$
that is linear in $\delta \bvec{a}$ (see figure~\ref{fig:Heff_Neff}).
This is a tensor with three indices, that reshaped in vector form, yields the
desired derivative $\nabla\rho_E(\bvec{a})=\vecmap(H_{eff}(\bvec{A}))$.
The computational cost for the \emph{exact} contraction of the tensor
networks in $H_{eff}(\bvec{A})$ scales as $O(ND^5)$~\cite{frank-2004a}.
We will give below a prescription of how this can be improved
\footnote{This is actually only an approximation of the exact contraction.
However, due to finite machine precision, there is effectively no
difference between both results.
}
to $O(mnD^3)+O(n^2D^3)$ for chains of arbitrary lengths.
Furthermore we will show how to choose the
smallest possible parameters $m$ and $n$ such that no loss in precision
occurs and why the scaling reduces to $O(mD^3)+O(nD^3)$ in the
case of very long chains.

The other piece that is necessary
for the computation of $\nabla E(\bvec{a})$, is the derivative occurring
in the second term of (\ref{eq:grad_Eexpval_vecA}); this term can be obtained in a
very similar way (see figure~\ref{fig:Heff_Neff}). We will use the notation
$N_{eff}(\bvec{A})$ for the object defined by
$\nabla\braket{\psi(\bvec{a})|\psi(\bvec{a})}
 =:\vecmap(N_{eff}(\bvec{A}))$.
Due to the simpler structure of the tensor network the computational cost
here will scale as $O(nD^3)$ for arbitrary chains and as $O(D^3)$
for very long chains.

Now let us introduce the following convention for denoting incomplete
tensor networks where merely one of the MPS tensors is missing:
$\langle O\rangle_\A^{[\overline{s}]}$ shall henceforth denote the
expectation value of the operator $O$ with respect to the TI MPS
defined by the tensor $\A$, where one tensor $\A$ has removed been
removed from $\ket{\psi(\A)}$ at site $s$.
Following this definition, the first term in the graphical
representation of $H_{eff}(\A)$ (see figure~\ref{fig:Heff_Neff})
reads $\langle H_{2,3}\rangle_\A^{[\overline{1}]}$.
If a tensor has been removed from $\bra{\psi(\A)}$ at site $s$, we
will denote this by underlining the site index, thus we write
$\langle O\rangle_\A^{[\underline{s}]}$. Using this convention
we can write $H_{eff}(\A)$ as

\begin{equation}\label{eq:Heff_00}
  H_{eff}(\A)=\sum_{s=1}^{N} \Big(
              \langle H_{1,2}\rangle^{[\overline{s}]}_\A+
              \langle H_{1,2}\rangle^{[\underline{s}]}_\A
              \Big)
  \,\,\,.
\end{equation}
\noindent
For real Hamiltonians and real MPS this reduces of course to

\begin{equation}\label{eq:Heff_01}
  H_{eff}(\A)=2\sum_{s=1}^{N}\langle H_{1,2}\rangle^{[\overline{s}]}_\A
  \,\,\,.
\end{equation}
\noindent
Similar considerations hold for $N_{eff}(\A)$. Thus, using $I$ to
denote the identity operator, we can rewrite the
gradient of the energy (\ref{eq:grad_Eexpval_vecA}) as

\begin{equation}\label{eq:grad_Eexpval_vecA_02}
  \nabla E(\a)=
  \frac{N H_{eff}(\A)}{\braket{\psi(\A)|\psi(\A)}}
 -\frac{N \rho_E(\a) N_{eff}(\A)}
       {\braket{\psi(\A)|\psi(\A)}^2}=
  2N \sum_{s=1}^{N} \Big(
  \frac{ \langle H_{1,2}\rangle^{[\overline{s}]}_\A }
       { \langle I\rangle_\A }
 -\frac{ \langle H_{1,2}\rangle_\A \langle I\rangle^{[\overline{s}]}_\A }
       { \langle I\rangle_\A^2 }
  \Big)
   \,\,\,.
\end{equation}

In the last part of this section we will briefly sketch how a gradient
based procedure can be employed to find ground states of PBC chains if one
is dealing with complex Hamiltonians and thereby complex MPS.
One possibility is to use a gradient
based algorithm that converges to a minimum of the real-valued function
$E:\mathbb{C}^{n}\rightarrow\mathbb{R}$ within the complex manifold
($n$ stands here for the number of independent complex
parameters in the MPS). It can be shown that in this case one obtains
the same expression (\ref{eq:grad_Eexpval_vecA_02}) for the gradient
of the energy albeit the individual terms are now complex valued vectors.
However, since standard library routines for gradient based
search cannot minimize over complex manifolds, let us mention
the second possibility just for the sake of completeness.
Due to $\a=\x+i\y$ with $\x,\y\in\mathbb{R}^n$,
one can treat the energy as an analytic function over a real manifold
with twice as many degrees of freedom, i.e.
$E:\mathbb{R}^{2n}\rightarrow\mathbb{R}$.
Similar considerations to the ones leading to
(\ref{eq:grad_Eexpval_vecA_02}) yield then for the gradient

\begin{equation}\label{eq:grad_Eexpval_vecA_03}
\begin{split}
   \nabla_{\x}E(\x,\y)&=
   \phantom{-}2N \sum_{s=1}^{N} \Big(
    \frac{ \Re\langle H_{1,2}\rangle^{[\overline{s}]}_\A }
         { \langle I\rangle_\A }
   -\frac{ \langle H_{1,2}\rangle_\A \Re\langle I\rangle^{[\overline{s}]}_\A}
         { \langle I\rangle_\A^2 } \Big) \\
   \nabla_{\y}E(\x,\y)&=
   -2N \sum_{s=1}^{N} \Big(
    \frac{ \Im\langle H_{1,2}\rangle^{[\overline{s}]}_\A }
         { \langle I\rangle_\A }
   -\frac{ \langle H_{1,2}\rangle_\A \Im\langle I\rangle^{[\overline{s}]}_\A}
         { \langle I\rangle_\A^2 } \Big)
   \,\,\,.
\end{split}
\end{equation}

\subsection{Computation of $H_{eff}(\A)$}

We introduce now a shorthand notation for the building blocks of
$H_{eff}(\bvec{A})$ that will allow us to express it in a very compact way.
From the graphical representation (see figure~\ref{fig:shorthand_defs}) it
should be obvious what the objects
$H_{AA}^{AA}$, $H_{AA}^{\phantom{A}A}$,
$H_{AA}^{A\phantom{A}}$, $T=T_A^A$ and $T_A$ mean;
note that $T$ denotes the MPS transfer matrix that has been
repeatedly mentioned in the previous sections.
For the sake of completeness we also give the definition of
the tensor $H_{AA}^{AA}$ explicitly in terms of its components:

\begin{figure}[ht]
  \begin{center}
    \includegraphics[width=0.9\textwidth]{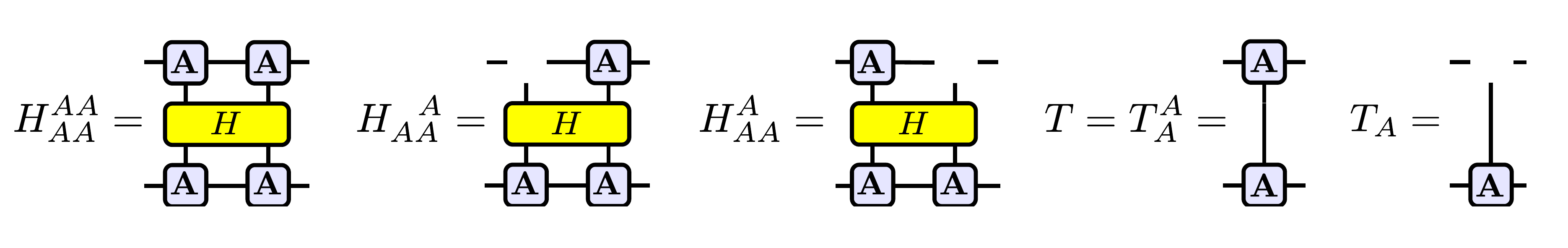}
  \end{center}
  \caption{
    (Color online). Graphical representation of $H_{AA}^{AA}$,
    $H_{AA}^{\hspace{2.3mm}A}$, $H_{AA}^{A}$, $T=T_A^A$ and $T_A$.
  }
\label{fig:shorthand_defs}
\end{figure}

\begin{equation}\label{eq:H_AAAA}
  (H_{AA}^{AA})^{\alpha\phantom{\alpha'}\gamma}
               _{\phantom{\alpha}\alpha'\phantom{\gamma}\gamma'}
  =A_{i\phantom{\alpha}\beta}^{\phantom{i}\alpha}
   A_{j\phantom{\beta}\gamma}^{\phantom{i}\beta}
   (H_{s,s+1})^{ij}_{\phantom{ij}i'j'}
   A_{\phantom{i'}\alpha'}^{i'\phantom{\alpha'}\beta'}
   A_{\phantom{j'}\beta'}^{j'\phantom{\beta'}\gamma'}
   \,\,\,.
\end{equation}
\noindent
Here we have used greek letters to label the virtual bonds,
latin ones for the physical bonds and Einstein summation convention to
denote contracted indices.
If one combines the left-hand side indices $\alpha$ and $\alpha'$ into
one big index and does the same for the right-hand side indices $\gamma$
and $\gamma'$, it is clear that $H_{AA}^{AA}$ represents a $D^2\times
D^2$-matrix.
The other objects defined in figure~\ref{fig:shorthand_defs} have similar
explicit definitions.
$H_{eff}(\A)$ now reads

\begin{equation}\label{eq:Heff_02}
  H_{eff}(\A)
 =2\cdot\tr^{*}\Big[
  H_{AA}^{\phantom{A}A} T^{N-2}+
  H_{AA}^{A\phantom{A}} T^{N-2}+
  \sum_{s=0}^{N-3} H_{AA}^{AA} T^s T_A T^{N-3-s}
  \Big]
\end{equation}
\noindent
where $\tr^{*}[\dots]$ indicates that the trace is taken only with respect
to the matrix multiplication of the "outer" indices of the "big"
$D^2\times D^2$-matrices. These "big" matrices may have internal open indices
that survive the $\tr^{*}[\dots]$-operation and make sure that
$H_{eff}(\A)$ is left with its tensor structure s.t. it can be later
reexpressed as a vector.

The computation of (\ref{eq:Heff_02}) is the bottleneck of our method.
If we would compute it by straightforward matrix multiplication, even  using the
sparseness, the computational cost would scale as $O(ND^5)$.
In order to improve this scaling, the crucial point is to realize that for
large $N$ most terms in (\ref{eq:Heff_02}) will contain high powers of $T$
which means that they can be very well approximated within the
subspace spanned by the dominant eigenvectors
\footnote{Normally one denotes the eigenvector corresponding to the
eigenvalue with the largest magnitude as the \emph{dominant eigenvector}.
Accordingly, the obvious meaning of the plural
(i.e. \emph{dominant eigenvectors}) would be to denote the eigenvectors
of a degenerated dominant eigenvalue. However, we rather use the term
\emph{dominant eigenvectors} in order to refer to a set of eigenvectors
whose corresponding eigenvalues have the largest magnitude among
all eigenvalues.
}
of $T$.
This can be easily seen if we write such factors in their
eigenbasis
\footnote{Obtaining this eigenbasis does not spoil the
overall computational cost as it scales better than the contractions of the
tensor networks. Due to the sparse structure of $T$, one can obtain its
$n$ dominant eigenvectors with $O(nD^3)$ operations.
}

\begin{equation}\label{eq:high_powers_transf_mat_01}
  T^s = \sum_{\alpha=1}^{D^2} \lambda_\alpha^s
        \ket{\lambda_\alpha}\bra{\lambda_\alpha}
      = \lambda_1^s \bigg[\ket{\lambda_1}\bra{\lambda_1}+
        \sum_{\alpha=2}^{D^2}
        \bigg(\frac{\lambda_\alpha}{\lambda_1}\bigg)^s
        \ket{\lambda_\alpha}\bra{\lambda_\alpha}\bigg]
\end{equation}
\noindent
where $\lvert\lambda_1\rvert \ge \lvert\lambda_2\rvert \dots
       \ge \lvert\lambda_{D^2}\rvert$.
Obviously the subspace corresponding to the small magnitude eigenvalues is
suppressed exponentially with $s$ and thus can be neglected for powers
$s$ that are large enough (e.g. for $s=20$ and
$\lvert \frac{\lambda_\alpha}{\lambda_1} \rvert\approx 0.1$,
$\lvert \frac{\lambda_\alpha}{\lambda_1} \rvert^s \approx10^{-20}<10^{-16}$
which is the machine precision of double precision floating point numbers).
In these cases it is perfectly fine to restrict ourselves to the subspace
spanned by say $n$ dominant eigenvectors, with the parameter $n$ yet to be
determined.
In fact, we will perform the entire computation a few times, starting
with a rather small $n$ and increasing it until the result does not improve
any more.
When this happens, we know that we have found the optimal $n$ beyond which,
when all other parameters are fixed, the precision does not get any better.
Thus we will approximate large powers of the transfer matrix as

\begin{equation}\label{eq:high_powers_transf_mat_02}
  T^s \approx \sum_{\alpha=1}^{n} \lambda_\alpha^s
        \ket{\lambda_\alpha}\bra{\lambda_\alpha}
        \,\,\,.
\end{equation}
\noindent
At this point we must remark that this approximation only works if the
moduli of the transfer matrix eigenvalues $\lvert\lambda_\alpha\rvert$
are not concentrated around a certain point (i.e. $T$ is not approximately
proportional to unity). In that case, any increment of
$n$ will improve the precision
and we will end up with very bad overall scaling
\footnote{In the extremal case of optimal $n=D^2$ the
overall scaling becomes $O(D^7).$
}.
For models where this behaviour occurs
the algorithm presented here may be worse than contracting the tensor
networks explicitly, where the scaling is $O(ND^5)$. In these cases the
chain length $N$ ultimately decides which method is preferable.
Fortunately for the models treated by us, this undesirable behaviour does not
occur and we end up with relatively small $n$ beyond which the precision
does not improve any more.

Let us now return to (\ref{eq:Heff_02}). There are two different types of terms
which must be treated differently. The first and the second term under our
somewhat unorthodoxly defined trace can be considered as "easy".
They are approximated by

\begin{equation}\label{eq:Heff_easy_projected}
  \langle H_{1,2}\rangle^{[\overline{1}]}_\A=
  \tr^{*}\Big[H_{AA}^{\phantom{A}A} T^{N-2}\Big] \approx \sum_{\alpha=1}^{n}
  \bra{\lambda_\alpha} H_{AA}^{\phantom{A}A}
  \ket{\lambda_\alpha} \lambda_\alpha^{N-2}
\end{equation}
\noindent
which is computed within $O(nD^3)$ operations. This is because each
contraction $\bra{\lambda_\alpha}H_{AA}^{\phantom{A}A}\ket{\lambda_\alpha}$
can be performed with cost $O(D^3)$
and this has to be done $n$ times.

The computationally more expensive terms are the ones under the sum over
$s$, where two different powers of $T$ are involved. We will call these
terms "hard". They are approximated by

\begin{equation}\label{eq:Heff_hard_projected}
  \langle H_{1,2}\rangle^{[\overline{3+s}]}_\A=
  \tr^{*}\Big[H_{AA}^{AA} T^s T_A T^{N-3-s}\Big]
  \approx \sum_{\alpha,\beta=1}^{n}
  \bra{\lambda_\beta} H_{AA}^{AA} \ket{\lambda_\alpha} \bra{\lambda_\alpha}
  T_A \ket{\lambda_\beta} \lambda_\alpha^{s} \lambda_\beta^{N-3-s}
  \,\,\,.
\end{equation}
\noindent
Here we must remark two things: \emph{i)} it is not necessary to let the second
index $\beta$ run over the same range as $\alpha$.
It would be possible to choose as an upper bound a further parameter
$n'$ and also vary this one until
the precision does not improve any more. However, since expression
(\ref{eq:Heff_hard_projected}) has obviously left-right
symmetry, it is sensible to assume that the optimal result would yield
$n=n'$. Even if this would not be the case, due to the fact that we scan along
$n$, convergence will be reached only for some
$n_{\textrm{optimal}} \ge \sup\{n,n'\}$, so we will find the lowest achievable
energy anyway;
\emph{ii)} for very small or very large $s$ either the left or the right
transfer matrix segments in (\ref{eq:Heff_hard_projected}) can not be
well approximated by a little number of eigenvalues $n$ since the
lower $\lambda_\alpha$ are not sufficiently suppressed by the small exponent.
In the worst case we would have to take all $D^2$ eigenvalues into
account, which dramatically increases the computational cost.
In order to solve this issue we will compute these terms by exact contraction
of segments of length $m$, which introduces this further parameter into our
algorithm. This will be explained in more detail further below.
For the moment let us note that depending on the magnitude of $s$,
we can further separate the sum in (\ref{eq:Heff_02}) over the "hard" terms
into

\begin{equation}\label{eq:Heff_hard_sum}
  \sum_{s=0}^{N-3}\equiv\sum_{s=0}^{m-1}+\sum_{s=m}^{N-3-m}+\sum_{s=N-2-m}^{N-3}
  \,\,\,.
\end{equation}
\noindent
We call the terms over which the second sum is taken "medium-$s$" terms
and will treat them differently from the "extremal-$s$" terms that appear
in the first respectively third sum.
Thus $H_{eff}(\A)$ can be divided into

\begin{equation}\label{eq:Heff_03}
  H_{eff}(\A)=2\cdot\bigg(H_{eff}^{easy}(\A)+
              H_{eff}^{hard,extr}(\A)+H_{eff}^{hard,med}(\A)
              \bigg)
  \,\,\,.
\end{equation}
\noindent

\subsubsection{Computation of "extremal-$s$" terms}

\begin{figure}[ht]
  \begin{center}
    \includegraphics[width=0.95\textwidth]{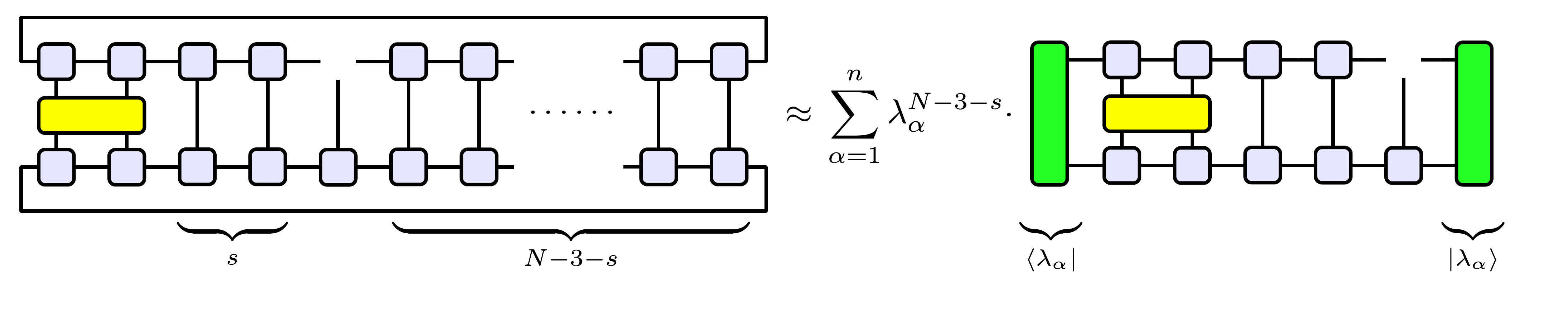}
  \end{center}
  \caption{
    (Color online). Graphical representation of a term with small
    $s$ and its approximation within the subspace spanned by
    $n$ dominant eigenvectors of $T$.
  }
\label{fig:Heff_extremal_s}
\end{figure}

In this section we treat the terms with small respectively large $s$.
The first thing to remark is that for large $N$, if $T^s$ can not
be well approximated within some low-dimensional subspace
because $s$ is too small, it is very likely that for $T^{N-3-s}$ the
approximation will work due to $N-3-s\gg s$.
The same observation holds in the other direction if $s$ is too large.
Secondly, depending on the
MPS bond dimension $D$ and the ammount of entanglement present in the MPS
(i.e. depending on the model one is treating),
there is a certain $m$ above
which $T^s$ with $s\geq m$ can be faithfully approximated within the
$n<D^2$-dimensional subspace spanned by $n$ dominant eigenvectors.
As we don't know anything about $m$ a priori,
we introduce it as a further parameter into our algorithm.
We will scan $m$ within its range $[1,1/2(N-2)]$ and in the end we will
obtain some optimal pair $(m,n)$.
The reason why
$m$ does not go all the way up to $N-3$ is that in order for our algorithm
to scale effectively as $D^3$, we must employ the dominant eigenvector
approximation on the other half of the chain.
Without it we would get the undesirable
scaling $O(ND^5)$. The contraction (see figure~\ref{fig:Heff_extremal_s})
we must perform for each term with small $s$ thus reads

\begin{equation}\label{eq:Heff_hard_small_s}
  \langle H_{1,2}\rangle^{[\overline{3+s}]}_\A=
  \tr^{*}\Big[H_{AA}^{AA} T^s T_A T^{N-3-s}\Big] \approx
  \sum_{\alpha=1}^{n} \bra{\lambda_\alpha} H_{AA}^{AA} T^s
  T_A \ket{\lambda_\alpha} \lambda_\alpha^{N-3-s}
  \,\,\,,\,\,\,\forall s<m
\end{equation}
\noindent
and can be done with computational cost $O(nD^3)$ using a sparse matrix
contraction scheme.
As we have to repeat this procedure $m$ times, the total cost scales as
$O(mnD^3)$.

The large $s$ terms (i.e. when $N-3-m<s\leq N-3$) can be easily obtained by
making use of the left-right symmetry of the tensor network around the point
with $s=(N-2)/2$. The sum over all these $s$ turns out to be related to
the sum over the small $s$ terms by taking
the transpose with respect to the
open virtual bond indices at the empty site where $T_A$ sits.
Thus the computational cost remains unchanged $O(mnD^3)$.

\subsubsection{Computation of "medium-$s$" terms}

For terms where $s$ is neither too small nor too large,
both powers of the transfer matrix (i.e. $T^s$ and $T^{N-3-s}$)
can be well approximated whithin the subspace
spanned by $n$ dominant eigenvectors.
The good news is that in this case the sum over $s$ can be performed
analitically in contrast to the "extremal-$s$" case where we had to
compute each of the $m$ terms separately. However, there is also bad
news, namely that we now have an additional sum over the eigenvalue
index stemming from the approximation of $T^{N-3-s}$.
Explicitly the sum over all "medium-$s$" reads

\begin{equation}\label{eq:Heff_hard_medium_s}
\begin{split}
   H_{eff}^{hard,med}(\A)&=
   \tr^{*}\Big[ \sum_{s=m}^{N-3-m} H_{AA}^{AA} T^s T_A T^{N-3-s} \Big]\\
 &\approx\tr^{*}\Big[ \sum_{s=0}^{N-3-2m} \sum_{\alpha,\beta=1}^{n}
    H_{AA}^{AA} T^m
    \ket{\lambda_\alpha} \lambda_\alpha^s \bra{\lambda_\alpha}
    T_A \ket{\lambda_\beta} \lambda_\beta^{N-3-2m-s} \bra{\lambda_\beta} T^m
    \Big]\\
 &=\sum_{\alpha,\beta=1}^{n}
    \bra{\lambda_\beta} H_{AA}^{AA}
    \ket{\lambda_\alpha} \bra{\lambda_\alpha}
    T_A \ket{\lambda_\beta} \lambda_\alpha^m \lambda_\beta^m
    \frac{\lambda_\beta^{N-2-2m} - \lambda_\alpha^{N-2-2m}}
         {\lambda_\beta - \lambda_\alpha}
   \,\,\,.
\end{split}
\end{equation}
\noindent
In the first step we have shifted the summation variable $s$ and have
written the matrices $T$ in their eigenbasis.
To arrive from the second to the third line
we have used the cyclic property of the trace to write the
entire expression as a sum over products of scalars (actually the factor
containing $T_A$ is only a scalar with respect to our specially defined
trace since it contains internal free indices).
Furthermore we have performed the $s$-sum straightforwardly.

The computational cost scales here as $O(n^2 D^3)$. This is because
we have two sums going from $1$ to $n$ over terms that are contracted
within $O(D^3)$ operations.

\subsection{Computation of $N_{eff}(\A)$}

Our prescription for the computation of $N_{eff}(\A)$ is also based on the
observation that big powers of the transfer matrix $T$ can be very well
approximated within the subspace spanned by the dominant eigenvectors. However
here things are much easier than for $H_{eff}(\A)$. This is because the
translational invariance is not broken by the 2-site Hamiltonian
(see figure~\ref{fig:Heff_Neff}) and we can write

\begin{equation}\label{eq:Neff_01}
  N_{eff}(\A)=2N\cdot \langle I\rangle^{[\overline{1}]}_\A
  \,\,\,.
\end{equation}
\noindent
Similarly to $\langle H_{1,2}\rangle^{[\overline{1}]}_\A$ in
(\ref{eq:Heff_easy_projected}), $\langle I\rangle^{[\overline{1}]}_\A$
is approximated by

\begin{equation}\label{eq:Neff_projected}
  \langle I\rangle^{[\overline{1}]}_\A=
  \tr^{*}\Big[T_{A} T^{N-1}\Big] \approx \sum_{\alpha=1}^{n}
  \bra{\lambda_\alpha} T_{A}
  \ket{\lambda_\alpha} \lambda_\alpha^{N-1}
\end{equation}
\noindent
which is computed within $O(nD^3)$ operations.

\subsection{Overall scaling of the computational cost}\label{sec:comp_cost}

We have seen that the gradients in (\ref{eq:grad_Eexpval_vecA})
can be obtained within $O(mnD^3)+O(n^2D^3)$ respectively $O(nD^3)$
operations if our approximation of large powers of the transfer matrix
is justified. It is easy to check that the scalar expectation values in
(\ref{eq:grad_Eexpval_vecA}) can be obtained in an analogue yet simpler way.
The fact that there are no vacant sites in the corresponding tensor
networks enables us to use everywhere a method identical to the
one used for $N_{eff}(\A)$. Thus the computational cost for our algorithm
scales as its most expensive part, namely as $O(mnD^3)+O(n^2D^3)$.

It is also not difficult to check that for very large chains
(i.e. either when $N \gg \xi$ for non-critical systems or
$N \gg \tilde{\xi}_{D}$ for critical ones, where $\tilde{\xi}_{D}$
is the effective correlation length induced by finite $D$)
this scaling
can be improved. First recall that we had in every tensor network at least
one portion of the chain expressed as a power of $T$ that we approximated
using its dominant eigenvectors.
Now, for any bond dimension $D$ there exists an $N$ above which all approximated
portions are long enough s.t. all eigenvalues except the largest one
are suppressed by the very large exponent. In this case the overall scaling
is $O(mD^3)+O(nD^3)$.
Note that in the scaling for the "extremal-s" terms we can not get rid of
$m$ because there will always be short portions between the $H_{1,2}$ and
the vacant site, that must be contracted exactly.
Similarly, for the "medium-s" terms (\ref{eq:Heff_hard_medium_s})
only the combinations of $\lambda_\alpha^m \lambda_\beta^m$
where both $\alpha$ and $\beta$ are large will be negligible.
Factors like $\lambda_1^m \lambda_\beta^m$ must usually always be
taken into account.
In any case, the ultimate check whether our approximations are justified
must be done in the simulations, where one must verify if there exists an
$n$ beyond which the ground state
\footnote{I.e. the state with the lowest energy which
we can achieve within the constrained
MPS-approximation that is used.
}
energy does not decrease.

We would like to compare our scaling of the computational cost to the
one of~\cite{pippan-2010} once again. Note that expressed
in the terms used in this work, the scaling from Ref.~\cite{pippan-2010} is
$O(NnD^3)$. On one hand,
as previously mentioned, our TI algorithm yields an improvement of one factor
$N$. On the other hand there is an additional factor $n$ that appears
in our scaling. This is due to the fact that we compute the gradient
of the energy explicitly.
It is easy to see that the computational cost for the evaluation of the
energy itself is $O(nD^3)$. However if we would restrict ourselves to
evaluations of the energy only, we would have to use something like a
downhill simplex method as the
outer function that scans the MPS manifold for the energy minimum.
In this case the outer function would call the energy evaluator
a huge number of times, thereby yielding the overall cost much higher
than one factor of $n$ that we must pay when computing the gradient.

\section{Numerical results}
\label{sec:numerical_results}

We have studied both critical and non-critical nearest neighbour
interaction spin models. The first one is the
Quantum Ising model for spins-$1/2$

\begin{equation}\label{eq:H_IS}
  H_{IS} = -\sum_{<ij>} Z_i Z_j - B\sum_{i}X_i
\end{equation}
\noindent
which we have simulated at its critical point $B=1$.
The second one is the antiferromagnetic Heisenberg model

\begin{equation}\label{eq:H_HB}
  H_{HB} = \frac{1}{2} \sum_{<ij>} (X_i X_j + Y_i Y_j + Z_i Z_j)\,\,\,.
\end{equation}
\noindent
This model is critical for spin-$1/2$ chains but non-critical for spin-$1$
chains. We have studied both cases.
Note that (\ref{eq:H_HB}) is not very well suited
for the description with 1-site TI MPS due to its antiferromagnetic
character. In order to cure this problem we apply in the case of the
spin-$1/2$ chain a global unitary
consisting of Pauli-$Y$ matrices on each second site
\footnote{
For the spin-$1$ chain we must apply the operator $M=\exp(i\pi Y)$ on
every second site in order to obtain the same effect.
}.
This leaves the spectrum unchanged and after we have found the 1-site TI
MPS for the ground state, we can recover the one for the unchanged
Hamiltonian by a new application of the global unitary.
The resulting MPS is then of course
2-site TI. The rotated Heisenberg Hamiltonian reads

\begin{equation}\label{eq:H_HBrot}
  H_{HB} = \frac{1}{2} \sum_{<ij>} (-X_i X_j + Y_i Y_j - Z_i Z_j)
  \,\,\,.
\end{equation}
\noindent

\subsection{Critical systems}
\label{sec:critical_systems}

\begin{figure}[ht]
  \begin{center}
    \includegraphics[width=1.0\textwidth]
    {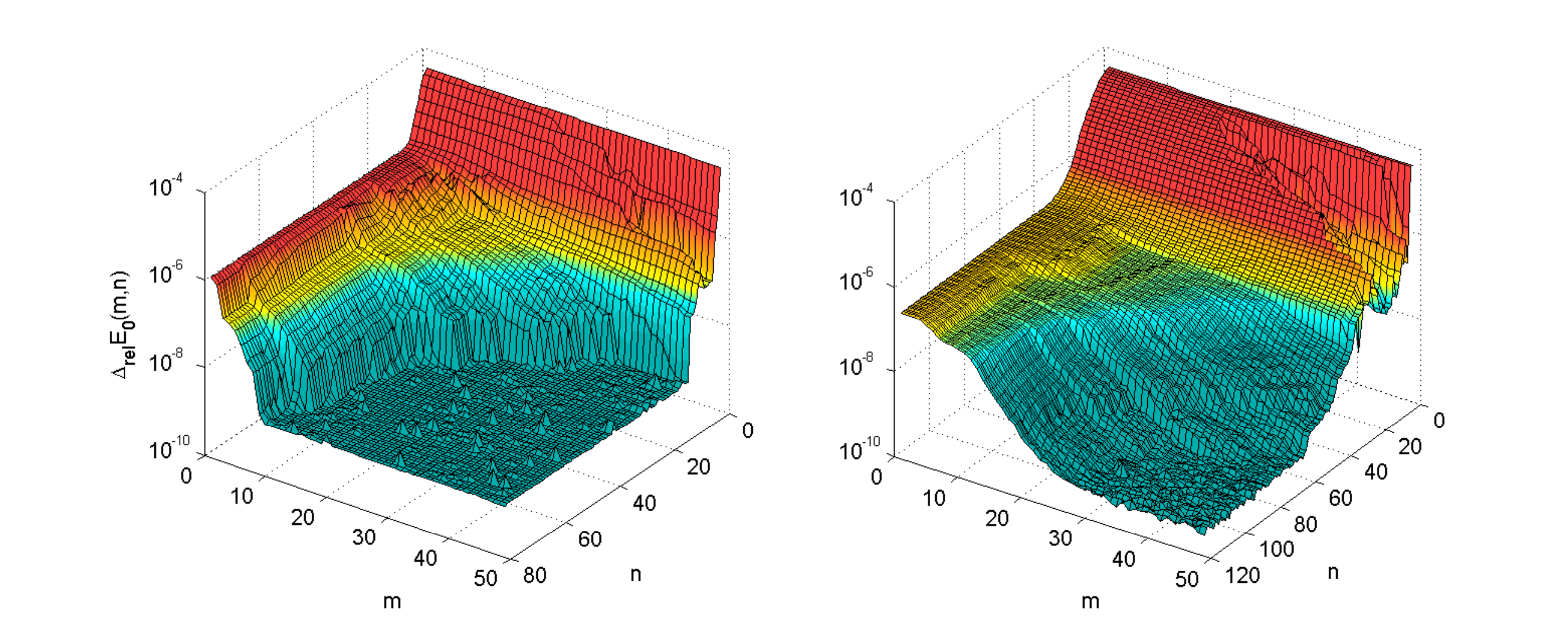}
  \end{center}
  \caption{
    (Color online).
    Critical Quantum Ising chain with $N=100$:
    Relative precision of the MPS ground state energy as compared
    to the analytical result as a function of the parameters $(m,n)$
    for $D=16$ (left) and $D=32$ (right).
  }
\label{fig:IS_N100_D16D32_mnscan_3D}
\end{figure}

\begin{figure}[ht]
  \begin{center}
    \includegraphics[width=1.0\textwidth]
    {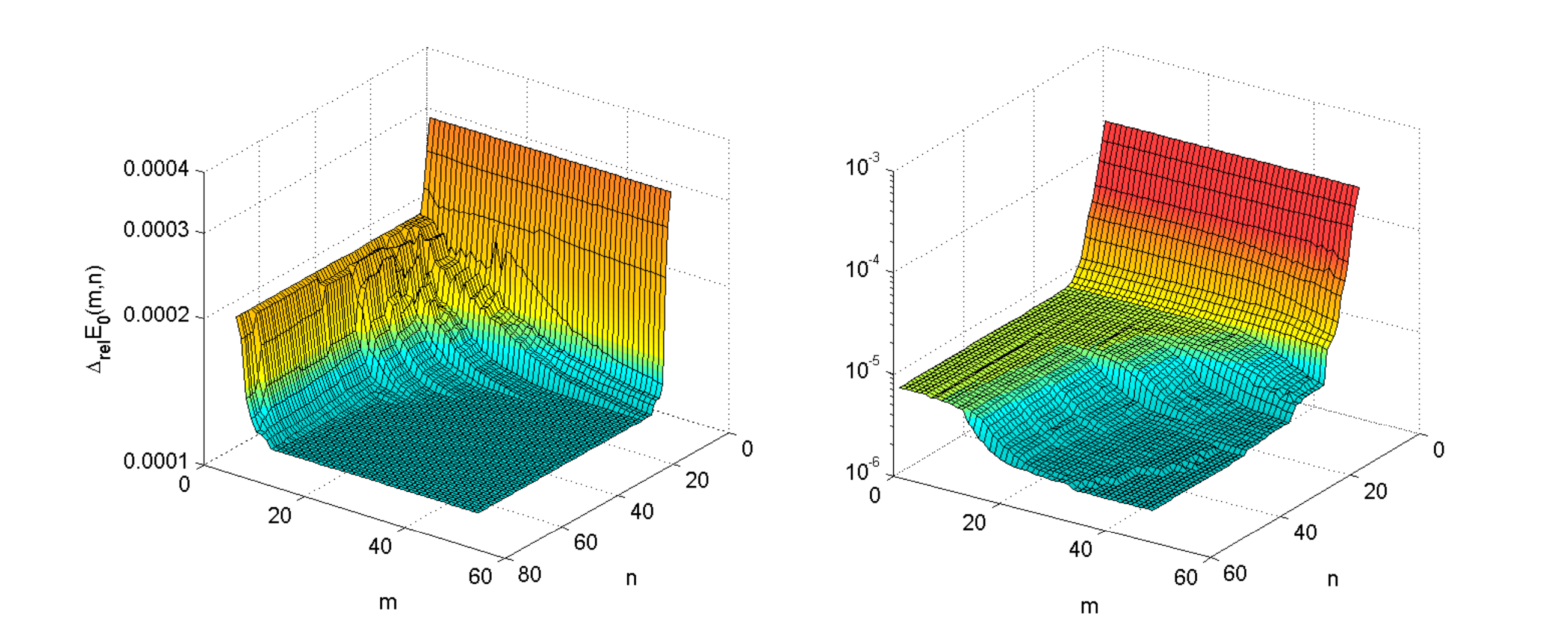}
  \end{center}
  \caption{
    (Color online).
    Critical Heisenberg chain with $N=100$:
    Relative precision of the MPS ground state energy as compared
    to the analytical result as a function of the parameters $(m,n)$
    for $D=16$ (left) and $D=32$ (right).
  }
\label{fig:HB_N100_D16D32_mnscan_3D}
\end{figure}

Let us illustrate the strategy for the scan of the parameter space spanned
by $\{m,n\}$ on the basis of results obtained for small critical
chains of $100$ and $400$ sites.
Figure~\ref{fig:IS_N100_D16D32_mnscan_3D} and
figure~\ref{fig:HB_N100_D16D32_mnscan_3D} show the relative precision
$\Delta_{rel}E_0(m,n)=(E_0^{exact}-E_0^{MPS}(m,n))/E_0^{exact}$ of
the MPS ground state energy compared to the exact solution as a function
of the algorithm parameters $m$ and $n$ for the
Quantum Ising respectively Heisenberg chain.
The first observation is that there exist $m_{max}$ and $n_{max}$
s.t. for all $m\ge m_{max}$, $n\ge n_{max}$ the precision does not
improve any more.
In the featured plots the plateau $\mathcal{P}$ with minimal energy
is reached within the plot range.
The optimal point $\{m_{opt},n_{opt}\}$ is then the point of
$\mathcal{P}$ that minimizes the scaling of the computational
cost $O(mnD^3)+O(n^2D^3)$ i.e.
$\{m_{opt},n_{opt}\}=\min\vert_{\{m,n\}\in\mathcal{P}}(mn+n^2)$.
Clearly, the optimal parameters $m_{opt}$ and $n_{opt}$ will be different
for different models and different values of the chain length $N$ and
the MPS bond dimension $D$.

The plots reveal a further detail: if we are not very pedantic about
the optimal $\{m,n\}$-pair, it is not necessary to scan the entire
plane, which is computationally very expensive.
If we are willing to settle for any pair $\{m, n\}$ that yields maximal
precision, we can scan along any line $n=km$ and we can be sure that at
some point we will hit $\mathcal{P}$.
This pair is quasi-optimal in the sense that we have
found the optimal $n$ for the corresponding $m$ and vice versa.
This is due to the fact that for any point of $\mathcal{P}$, especially
for its boundary, walking along lines with increasing $m$ or $n$ does
not take us out of $\mathcal{P}$.
As one can see in figure~\ref{fig:IS_N100_D16D32_mnscan_3D} and
\ref{fig:HB_N100_D16D32_mnscan_3D}, $\mathcal{P}$ is roughly
symmetric in $m$ and $n$, so a sensible line to scan along is given
by $n=m$
\footnote{In practice it might be better to choose $k<1$ since there
are parts of the algorithm with the scaling $O(n D^3)$ multiplied by
a big constant factor. In our simulations we have used $k=1/5$.
}.
As we have mentioned before, our algorithm
allows us to increase $m$ only up to $(N-2)/2$. If until then,
the results obtained along $n=m$ have not converged yet, we must continue
the scan along the line given by the constant maximal $m$ towards larger $n$.

\begin{figure}[ht]
  \begin{center}
    \includegraphics[width=1.0\textwidth]
    {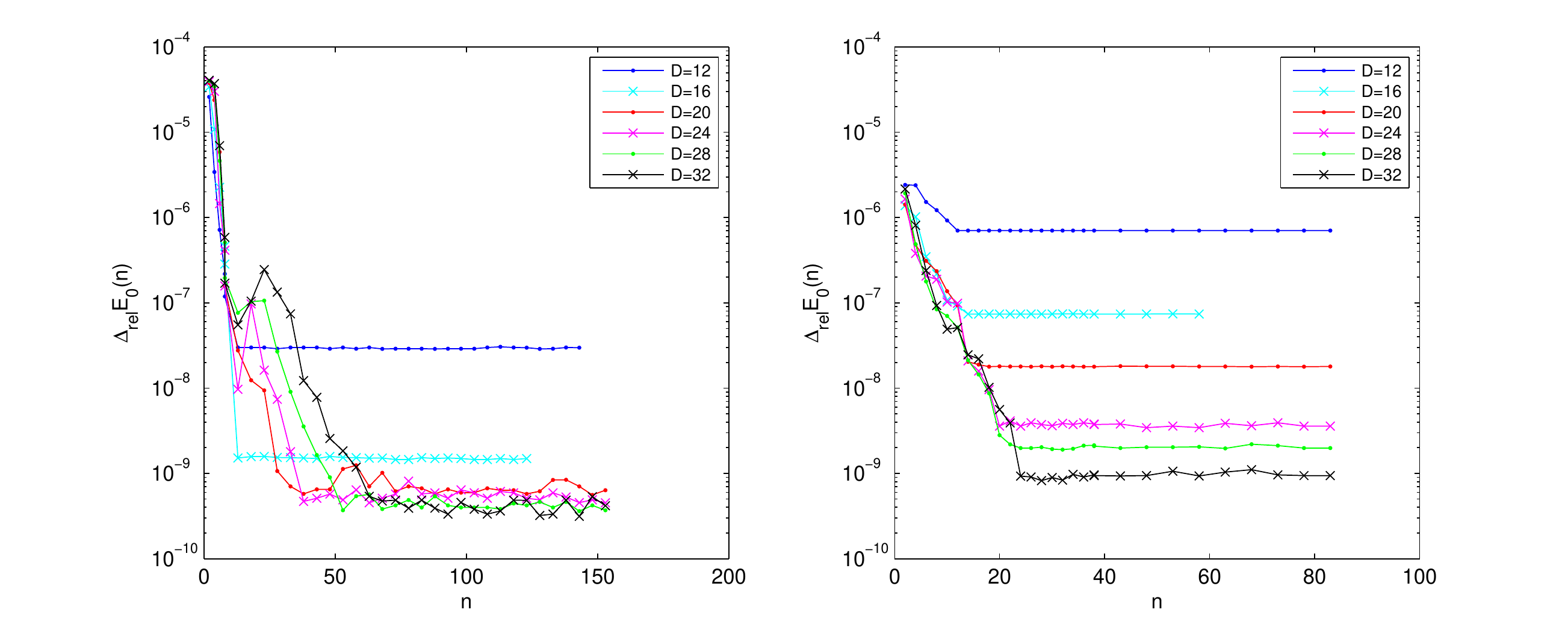}
  \end{center}
  \caption{
    (Color online).
    Critical Quantum Ising chain with $N=100$ (left) and $N=400$ (right):
    Relative precision of the MPS ground state energy
    as a function of the parameter $n$ for different bond dimensions $D$.
    The scan has been performed along the line $m=5n$ up to the maximal
    value of $m$ and then along the line with constant $m=(N-2)/2$.
  }
\label{fig:IS_N100N400_Dall_linescan}
\end{figure}

The relative precision of the MPS ground state energy for such line scans
is plotted in figure~\ref{fig:IS_N100N400_Dall_linescan}.
We notice that with increasing $D$ the maximally reachable precision
gets better in concordance to what one would expect.
The fact that $m_{opt}$ and $n_{opt}$ increase with $D$ is also intuitive.
What is a bit surprising is that for small $n$ the results obtained for
small bond dimensions are either similar or even better than the ones
obtained for higher bond dimensions.
This means that if one is not willing to go to larger values of $n$,
there is no point in increasing $D$!

Another interesting point is that for fixed $D$, as we increase $N$,
the plateau $\mathcal{P}$ is reached sooner and sooner
(i.e. for smaller values of $n$ and implicitly of $m$).
This behaviour is due to the fact that with increasing $N$ the
weight that we loose in our contracted tensor network by choosing
$n<D^2$ becomes negligible at smaller $n$.

\subsection{Observables - energy and correlation functions}

\begin{figure}[ht]
  \begin{center}
    \includegraphics[width=1.0\textwidth]
    {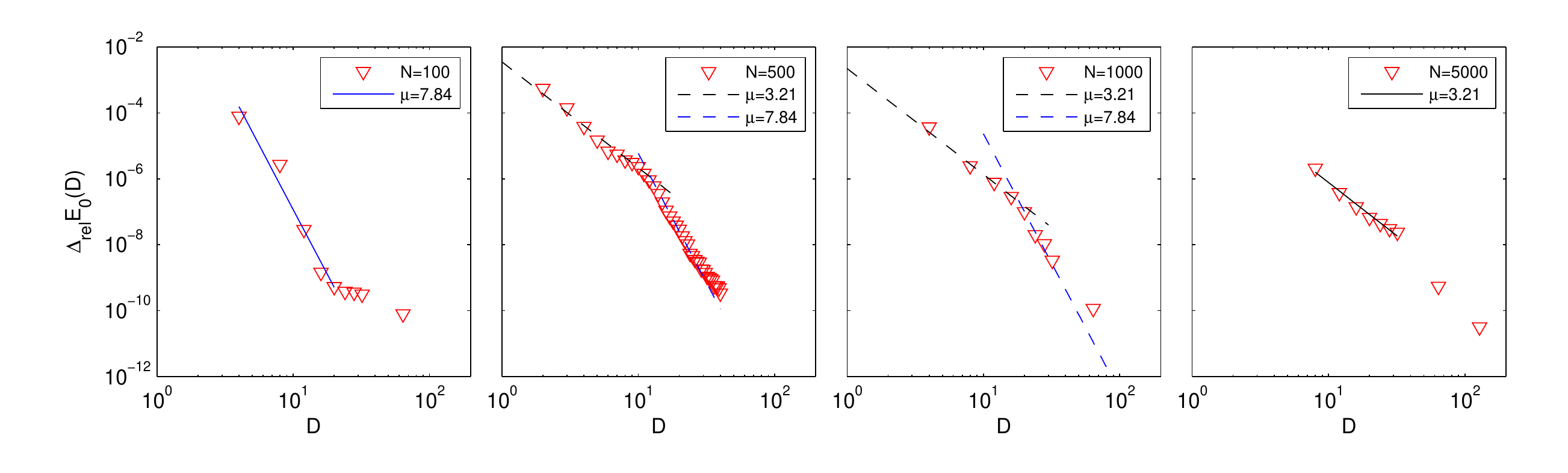}
  \end{center}
  \caption{
    (Color online).
    Critical Quantum Ising model:
    relative precision of the MPS ground state energy for
    different $N$ as a function of $D$.
  }
\label{fig:IS_N100N500N1000N5000_Dall_DE}
\end{figure}

\begin{figure}[ht]
  \begin{center}
    \includegraphics[width=1.0\textwidth]
    {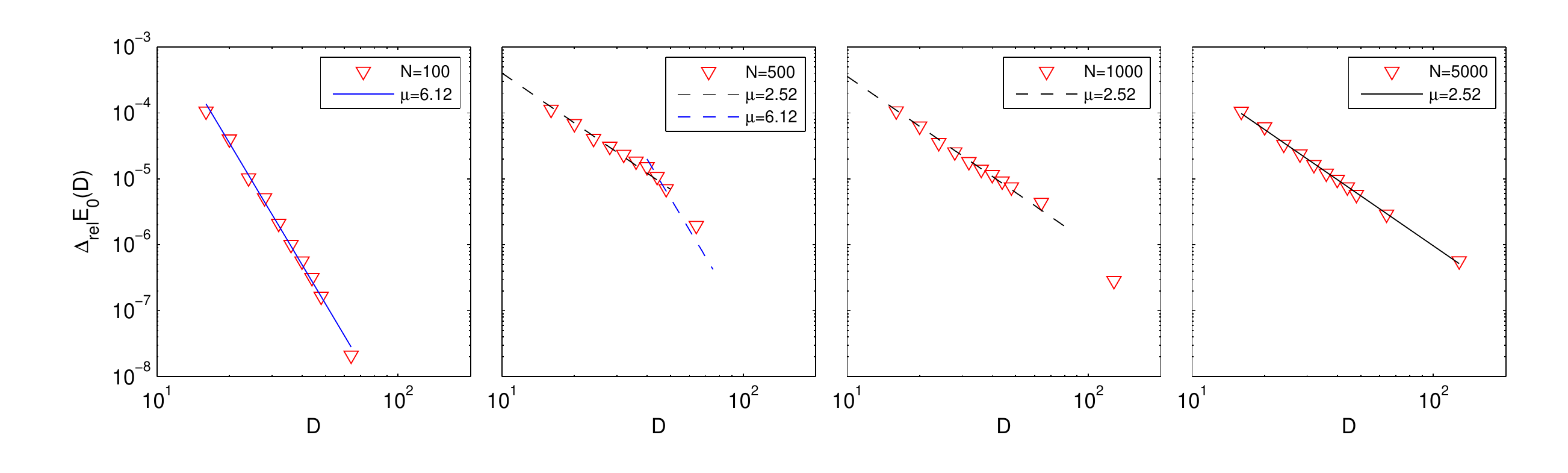}
  \end{center}
  \caption{
    (Color online).
    Critical Heisenberg model:
    relative precision of the MPS ground state energy for
    different $N$ as a function of $D$.
  }
\label{fig:HB_N100N500N1000N5000_Dall_DE}
\end{figure}

As the computational cost of our algorithm actually decreases if
we increase the number of sites $N$ while keeping $D$ constant,
we can investigate PBC chains of arbitrary size
\footnote{
However the precision is getting worse if we increase the
chain length without increasing $D$.
}.
Figure~\ref{fig:IS_N100N500N1000N5000_Dall_DE} and
figure~\ref{fig:HB_N100N500N1000N5000_Dall_DE} show the relative
precision of the ground state energy for the
critical Quantum Ising respectively Heisenberg model as a function
of the MPS bond dimension $D$.
We can see that generally the relative error is decreasing
as a polynomial of $D$ i.e. $\Delta_{rel}E_0(D)\propto D^{-\mu}$.
We have fitted straight lines through the reliable
\footnote{
If $D$ is too large for a given chain length $N$, the optimal parameter
$n$ can get close to its maximal value i.e. $n\approx D^2$. In these
cases the line scan described in section~\ref{sec:critical_systems}
converges at moderate $n$ only due to finite machine precision.
However, the precision of the MPS that is obtained in this way is
not the one that is theoretically maximally achievable with an MPS
of bond dimension $D$. We emphasize that with infinite
machine precision the line scan with converge only close to $n=D^2$
and also the large $D$ points in
figure~\ref{fig:IS_N100N500N1000N5000_Dall_DE} and
figure~\ref{fig:HB_N100N500N1000N5000_Dall_DE} would lie roughly on
the line corresponding to polynomial decay.
}
data of the $N=100$ and $N=5000$ plots and have obtained for
the exponent $\mu$ the values $7.84$ and $3.21$ ($6.12$ and $2.52$)
for the critical Quantum Ising (Heisenberg) model.
In the central plots (i.e. $N=500$ and $N=1000$) one can distinguish
between two regions where the relative precision is decaying polynomially
with the exponents obtained from the outer plots (i.e. $N=100$ and $N=5000$).
We have emphasized this by drawing dashed lines through the data points
in the central plots. Note that the dashed lines are not fitted, they have
merely the same slope as the full lines in the outer plots.
This behaviour can be best understood if one looks at
correlation functions.

\begin{figure}[ht]
  \begin{center}
    \includegraphics[width=1.0\textwidth]{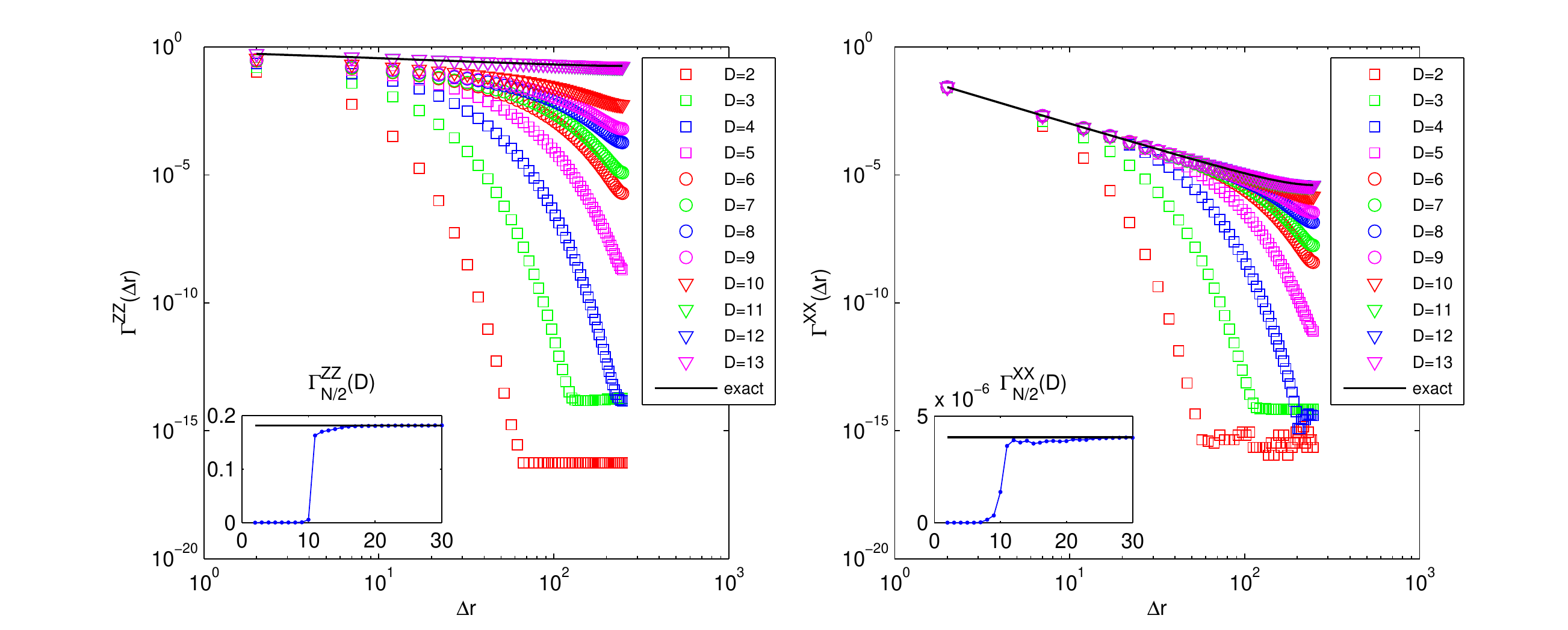}
  \end{center}
  \caption{
    (Color online).
    Correlation functions for a critical Quantum Ising chain with
    $N=500$. Left: order parameter correlator $\Gamma^{ZZ}(\Delta r)$
    and as inset the half-chain correlator as a function of $D$.
    Right: correlator $\Gamma^{XX}(\Delta r)$
    and as inset the half-chain correlator as a function of $D$.
  }
\label{fig:IS_N500_CF}
\end{figure}

Let us first consider the critical Quantum Ising model.
In figure~\ref{fig:IS_N500_CF} we have plotted the
$ZZ$ and the $XX$ correlation functions $\Gamma^{ZZ}(\Delta r)$ and
$\Gamma^{XX}(\Delta r)$
\footnote{
$\Gamma^{ZZ}(\Delta r)=\langle Z_r Z_{r+\Delta r}\rangle -
                       \langle Z_r \rangle \langle Z_{r+\Delta r}\rangle$,
$\Gamma^{XX}(\Delta r)=\langle X_r X_{r+\Delta r}\rangle -
                       \langle X_r \rangle \langle X_{r+\Delta r}\rangle$.
} in the MPS ground state of a chain with $N=500$ sites.
The solid line represents the exact solution obtained by applying
the programme of~\cite{lsm-1961} to the Quantum Ising model with PBC.
One can clearly see that with increasing $D$ the MPS correlations become
more and more accurate, just as one would expect. Note that we have
only plotted the correlation functions for separations
$\Delta r\le N/2$. This is because due to the periodic boundary
conditions $\Gamma(\Delta r)$ is symmetric around $\Delta r=N/2$
\footnote{This holds for even $N$. In the case of odd $N$ we have
$\Gamma((N-i)/2)=\Gamma((N+i)/2),\forall i\in\{1,3,5,\dots,N-2\}$.
}.
We would like to point out that while the exact $\Gamma(\Delta r)$ is linear
for small $\Delta r$ thus implying polynomial decay of
correlations in that regime, it flattens out towards $\Delta r\approx N/2$.
This behaviour is consistent with the physical requirement that the
correlation function is smooth at $\Delta r=N/2$.
The insets show the value of the half-chain correlators
$\Gamma_{N/2}(D):=\Gamma(\Delta r=N/2,D)$ as a function of $D$.
One can clearly see a jump in $\Gamma_{N/2}(D)$ at some $D'$.
This means that in this model, if one wants to obtain good
approximations for long range correlations in the ground state,
one must use MPS with bond dimension $D\ge D'(N)$.
Note that the jump in the inset of figure~\ref{fig:IS_N500_CF} occurs
roughly in the same region as the change of the slope in the second
plot of figure~\ref{fig:IS_N100N500N1000N5000_Dall_DE}.
This allows us to understand why in
figure~\ref{fig:IS_N100N500N1000N5000_Dall_DE}
the slope for large $D$ is steeper
than the one for small $D$: if $D$ is not large enough such that correlations
are faithfully reproduced throughout the entire chain, this represents
a further source of error besides the inherent error of MPS with
non-exponential bond dimension (i.e. $D\ll d^{N/2}$).

\begin{figure}[ht]
  \begin{center}
    \includegraphics[width=1.0\textwidth]{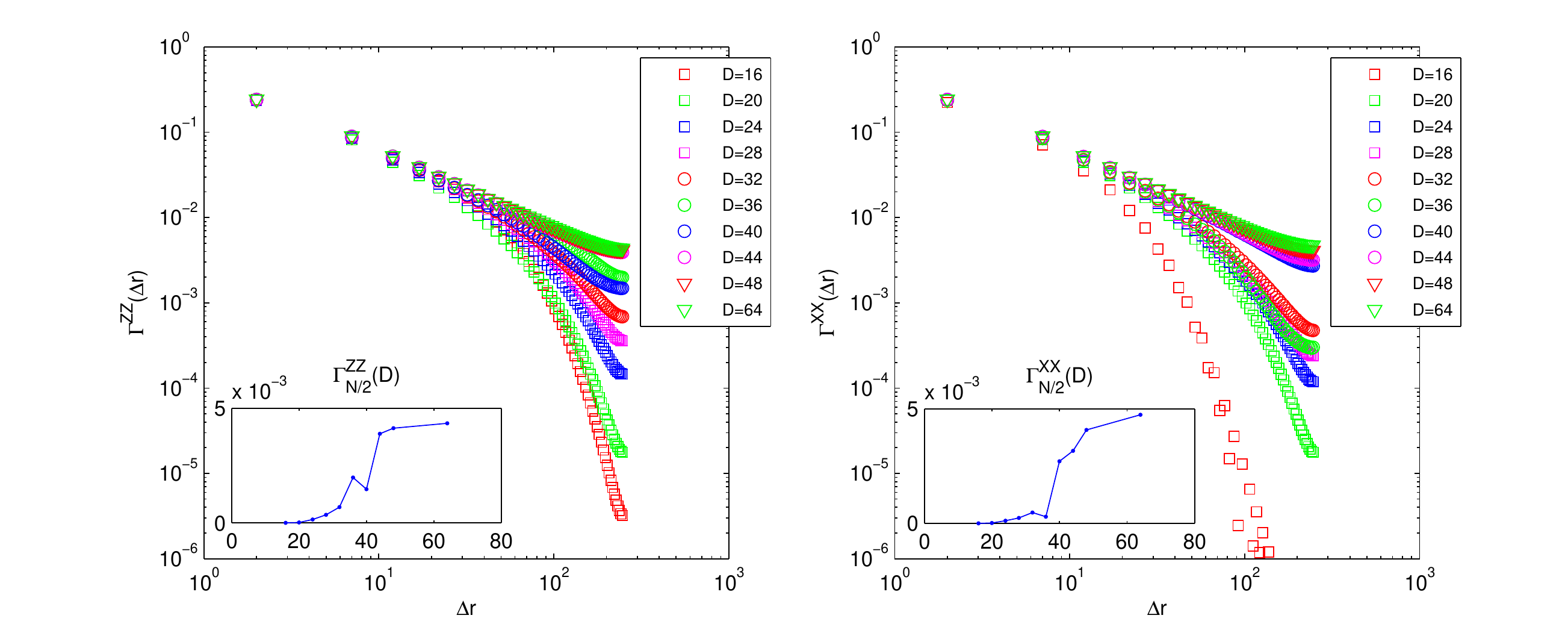}
  \end{center}
  \caption{
    (Color online).
    Absolute value of the correlation functions for a critical Heisenberg
    chain with $N=500$. Left: correlator $\Gamma^{ZZ}(\Delta r)$
    and as inset the half-chain correlator as a function of $D$.
    Right: correlator $\Gamma^{XX}(\Delta r)$
    and as inset the half-chain correlator as a function of $D$.
  }
\label{fig:HB_N500_CF}
\end{figure}

The absolute value of the correlation functions
\footnote{
Due to the antiferromagnetic nature of the Heisenberg model
the groundstate correlation function is changing its sign from
site to site.
}
for the critical Heisenberg chain with $N=500$
sites can be found in figure~\ref{fig:HB_N500_CF}. Note that these plots
only contain the MPS data since we do not have analytical expressions
for the long range correlations. Qualitatively figure~\ref{fig:HB_N500_CF}
shows the same behaviour as figure~\ref{fig:IS_N500_CF}. Quantitatively
we can see that correlation functions converge at much larger $D$ than in
the case of the critical Quantum Ising model, which is exactly what we
would expect. The half-chain correlators $\Gamma_{N/2}(D)$ exhibit
a more or less continuous transition to the region where correlations
are faithfully reproduced.

\begin{figure}[ht]
  \begin{center}
    \includegraphics[width=1.0\textwidth]{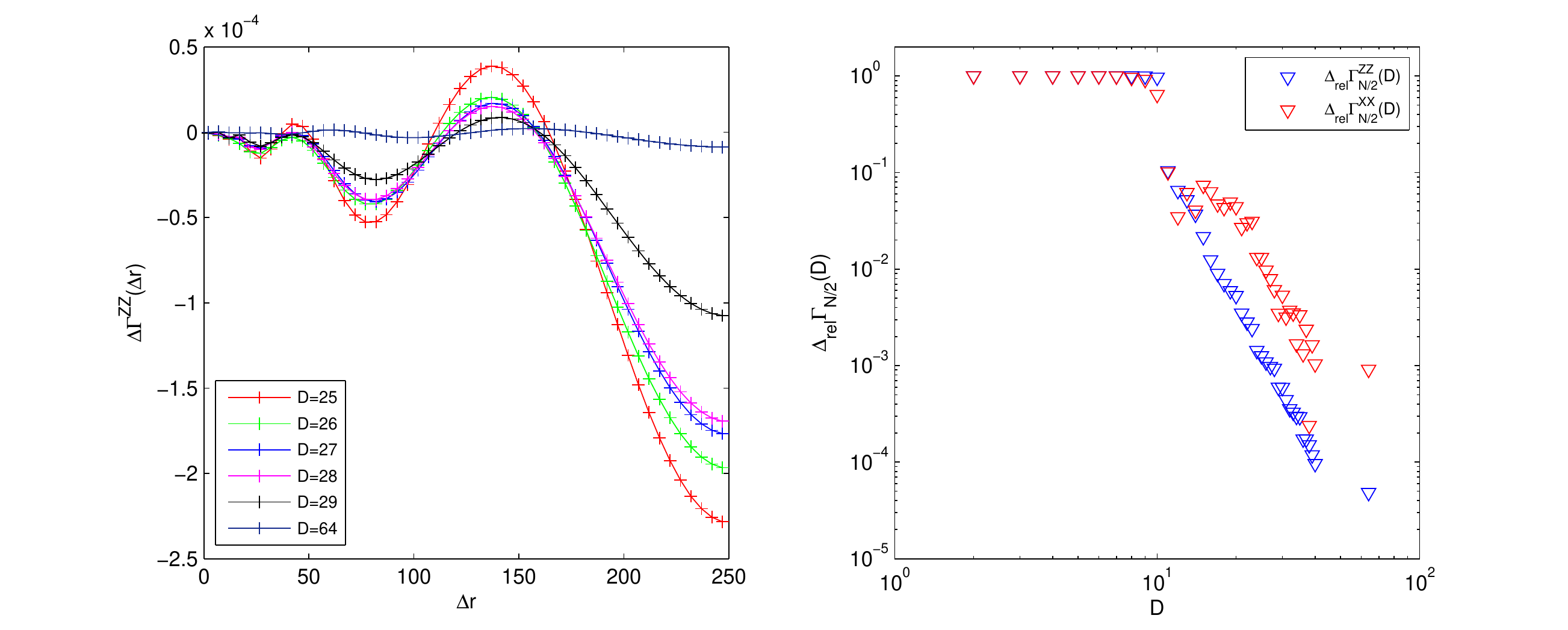}
  \end{center}
  \caption{
    (Color online).
    Relative precision of correlation functions in the MPS ground state
    of the Quantum Ising model with $N=500$.
    Left: error of the order parameter correlation
    function $\Gamma^{ZZ}(\Delta r)$ for several different $D$
    in the high precision regime.
    Right: relative precision of the half-chain
    correlators $\Gamma_{N/2}^{ZZ}$ and $\Gamma_{N/2}^{XX}$
    as a function of the MPS bond dimension $D$.
  }
\label{fig:IS_N500_DELTArelCF}
\end{figure}

We would like to make an interesting final remark regarding the error
in the correlation functions as a function of $\Delta r$.
In the left part of figure~\ref{fig:IS_N500_DELTArelCF}
we have plotted $\Gamma^{ZZ}_{MPS}(\Delta r)-\Gamma^{ZZ}_{exact}(\Delta r)$
for different $D$ in the regime where the half-chain correlators have
well converged (i.e. $D>25$).
The surprising thing is that the error does not grow
monotonically as a function of $\Delta r$ as one would expect,
but that it rather oscillates around zero. Nevertheless the amplitude
of the oscillations is growing monotonically with $\Delta r$.
The right part of figure~\ref{fig:IS_N500_DELTArelCF} reveals
that similary to the relative error of the ground state energy,
the relative error of the half-chain correlators
$\Delta_{rel}\Gamma_{N/2}(D)$ obeys power-law decay as a function of $D$
in the large $D$ regime.

Our numerical analysis thus indicates that for each $N$ there is a minimum
value of $D=D'(N)$ such that correlations throughout the entire chain
are properly captured. As investigated in~\cite{inprep-2010},
for critical systems this minimum value of $D'(N)$ is seen to be given by
a small power of $N$ that depends on the universality class of the model.
This dependence will allow us in~\cite{inprep-2010}
to characterize the cost of the algorithm presented in this work as
a power of $N$.
For the moment we will settle for a scaling of the 
overall computational cost of $O(g(D,\xi/N)D^3)$
where $g(D,\xi/N)$ will be seen to become trivial only for
non-critical systems.

\subsection{Non-critical systems}
\label{sec:noncritical_systems}

\begin{figure}[ht]
  \begin{center}
    \includegraphics[width=1.0\textwidth]
    {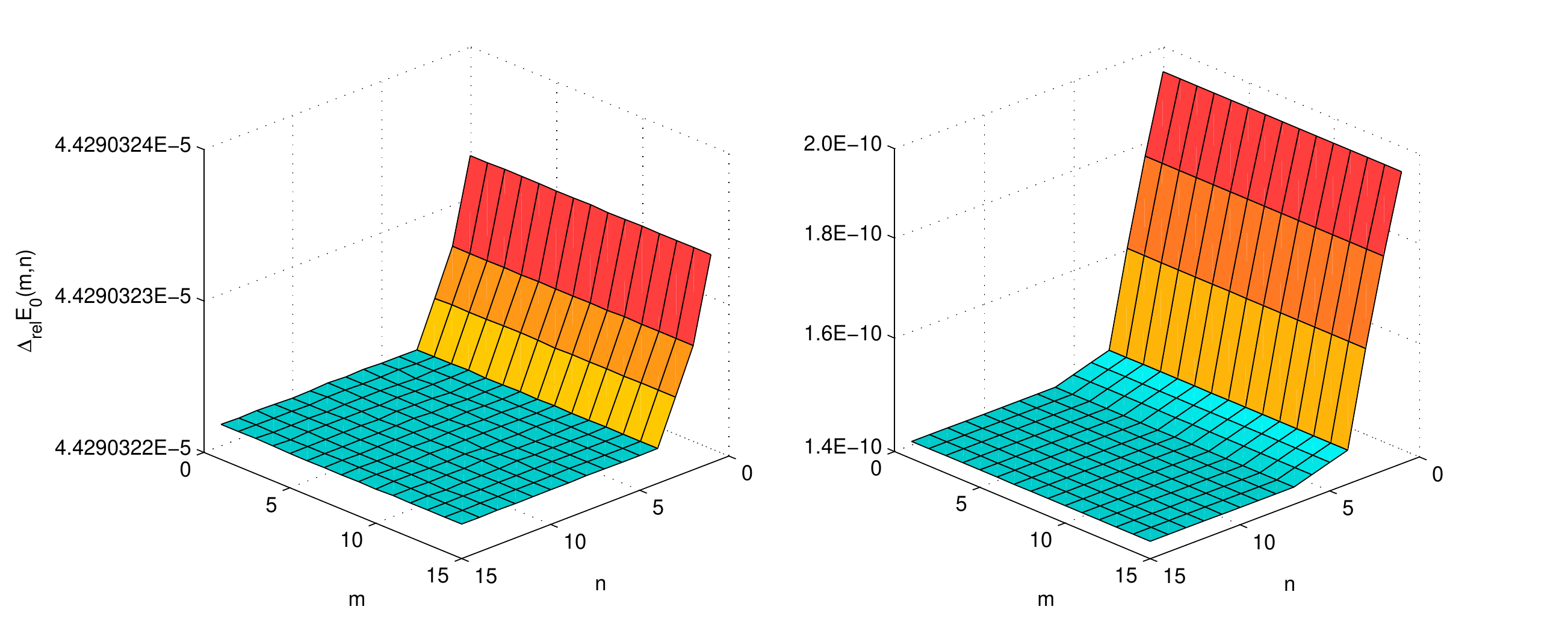}
  \end{center}
  \caption{
    (Color online).
    Spin-$1$ Heisenberg chain with $N=100$:
    Relative precision of the MPS ground state energy as compared
    to the best numerical approximation as a function of the parameters
    $(m,n)$ for $D=16$ (left) and $D=100$ (right).
  }
\label{fig:HBS1_D16D100_N100_DE_mnscan}
\end{figure}

We have seen that for critical systems it is quite involved
to predict the computational cost of MPS algorithms that find
the optimal approximation of the ground state within the manifold
defined by MPS with fixed bond dimension $D$. This turns out to
be much easier for non-critical systems where the correlation length
$\xi$ is much smaller than the chain length $N$. We have studied
the spin-$1$ Heisenberg chain as the prototype of a non-critical
quantum spin chain in order to be able to compare our results with
the ones presented in Ref.~\cite{pippan-2010}. As pictured in
figure~\ref{fig:HBS1_D16D100_N100_DE_mnscan}, for $N=100$ and $D$ that
is not too big, $n=4$ is sufficient in order to obtain the optimal
MPS approximation to the ground state. This is in agreement with
the predictions of Ref.~\cite{pippan-2010}. However for $D$ as big as
$100$, we would have to choose $n=7$ if we are not willing to loose
any precision. This indicates a dependence of $n$ on $D$ which is much
weaker than in the case of critical systems. Since due to finite
computer memory we cannot increase $D$ arbitrarily, it is safe to
say that for systems where $\xi\ll N$, $n$ is given by a small constant.
This is exactly what happens for a spin-$1$ Heisenberg chain with
$100$ sites since as shown in Ref.~\cite{white-huse-1993} the correlation
length is roughly $\xi\approx 6$ s.t. $\xi\ll N$.
It is obvious from figure~\ref{fig:HBS1_D16D100_N100_DE_mnscan}
that $m$ can be chosen arbitrarily so we can fix it to $m=1$.
Thus in this case the cost of our algorithm scales like $O(D^3)$
which is indeed by a factor $N$ less than the cost from~\cite{pippan-2010}.
Nevertheless we must emphasize that for systems where the condition
$\xi\ll N$ is not fulfilled anymore, the picture of a small constant
$n$ breaks down and the characterization of the computational cost
becomes non-trivial.

\begin{figure}[ht]
  \begin{center}
    \includegraphics[width=1.0\textwidth]{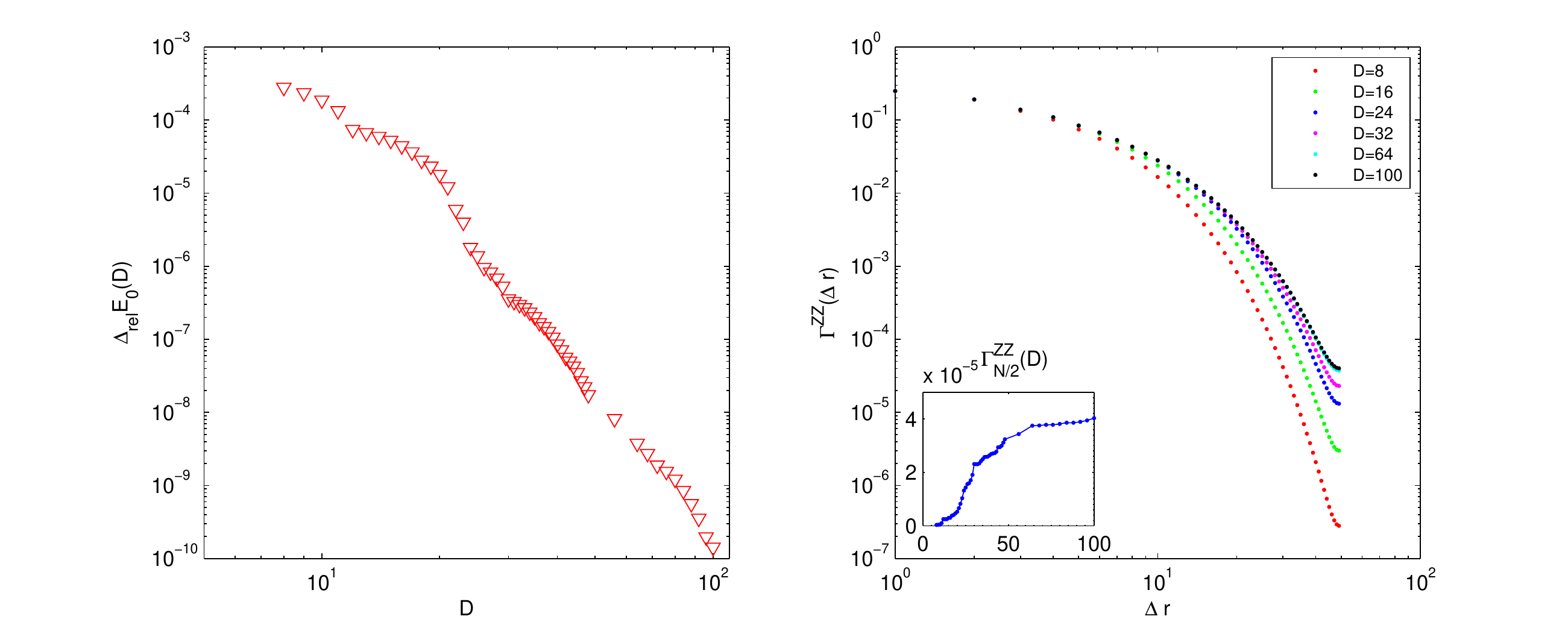}
  \end{center}
  \caption{
    (Color online).
    Spin-$1$ Heisenberg chain with $N=100$:
    Left: relative precision of the MPS ground state energy
    as a function of $D$.
    Right: absolute value of the correlation functions
    $\Gamma^{ZZ}(\Delta r)$
    and as inset the half-chain correlator as a function of $D$.
  }
\label{fig:HBS1_N100_DE_CF}
\end{figure}

In figure~\ref{fig:HBS1_N100_DE_CF} we have plotted the relative
energy precision and the correlation functions as functions of $D$.
Note that for the "exact" ground state energy density
we have used $E_0=-1.401484039$ which is the value obtained by
an extrapolation of our own finite $D$ results to infinite $D$.
We have done this since
the ground state energy that we obtain for $D=100$ is smaller than
any other value we have found in the literature, and in particular smaller
than the one used as the "exact" ground state energy in
Ref.~\cite{pippan-2010}.

The correlation functions plotted in figure~\ref{fig:HBS1_N100_DE_CF}
show non-trivial behaviour around $\Delta r\approx N/2$ where they
clearly deviate from exponential decay. The half-chain correlator
plotted in the inset seems to converge as a function of $D$ but we
do not have compelling evidence for that.


\section{Conclusions}
\label{sec:conclusions}

We have demonstrated the performance of a gradient based algorithm
for the simulation of TI spin chains with PBC both
for critical and non-critical systems.
For critical systems where the correlation length is of the order of
the system size, the overall scaling of the computational cost is
$O(mnD^3)+O(n^2D^3)$ and we
have given an analysis of the parameter space $\{m,n\}$ with a
prescription of how to obtain a quasi-optimal pair $\{m_{opt},n_{opt}\}$.
In the special case of a critical system that is simulated by
MPS with comparatively small $D$, such that 
$\tilde{\xi}_D\ll N$ holds for the induced correlation length,
the overall scaling is given by $O(mD^3)+O(nD^3)$.
For non-critical systems with a correlation length that is much smaller
than the system size, increasing $D$ barely affects the parameters
$m$ and $n$ and we can write for the overall scaling $O(D^3)$.
In the last two cases the cost is one
factor $N$ less than the one of the algorithm presented
in Ref.~\cite{pippan-2010}.
However, for critical systems in the large-$D$ regime,
the cost of Ref.~\cite{pippan-2010} is improved merely by a factor
$N/n$ due to the appearence of $n^2$ in the scaling of 
our algorithm. The precision of our numerical results
is comparable with or even better than that of previous algorithms
with the same bond dimension.
With a TI MPS approximation of the ground state at hand it is possible
to develop efficient MPS algorithms for the computation of excitations
in TI systems.

\section{Acknowledgements}

We thank V. Murg, E. Rico and L. Tagliacozzo for valuable discussions.
This work was supported by the
FWF doctoral program Complex Quantum Systems (W1210)
the FWF SFB project FoQuS, the ERC grant QUERG, and the ARC grants
FF0668731 and DP0878830.



\begin{mcitethebibliography}{24}
\expandafter\ifx\csname natexlab\endcsname\relax\def\natexlab#1{#1}\fi
\expandafter\ifx\csname bibnamefont\endcsname\relax
  \def\bibnamefont#1{#1}\fi
\expandafter\ifx\csname bibfnamefont\endcsname\relax
  \def\bibfnamefont#1{#1}\fi
\expandafter\ifx\csname citenamefont\endcsname\relax
  \def\citenamefont#1{#1}\fi
\expandafter\ifx\csname url\endcsname\relax
  \def\url#1{\texttt{#1}}\fi
\expandafter\ifx\csname urlprefix\endcsname\relax\def\urlprefix{URL }\fi
\providecommand{\bibinfo}[2]{#2}
\providecommand{\eprint}[2][]{\url{#2}}

\bibitem[{\citenamefont{White}(1992)}]{white-1992}
\bibinfo{author}{\bibfnamefont{S.~R.} \bibnamefont{White}},
  \bibinfo{journal}{Phys. Rev. Lett.} \textbf{\bibinfo{volume}{69}},
  \bibinfo{pages}{2863} (\bibinfo{year}{1992})\relax
\mciteBstWouldAddEndPuncttrue
\mciteSetBstMidEndSepPunct{\mcitedefaultmidpunct}
{\mcitedefaultendpunct}{\mcitedefaultseppunct}\relax
\EndOfBibitem
\bibitem[{\citenamefont{Rommer and \"Ostlund}(1997)}]{rommer-ostlund-1997}
\bibinfo{author}{\bibfnamefont{S.}~\bibnamefont{Rommer}} \bibnamefont{and}
  \bibinfo{author}{\bibfnamefont{S.}~\bibnamefont{\"Ostlund}},
  \bibinfo{journal}{Phys. Rev. B} \textbf{\bibinfo{volume}{55}},
  \bibinfo{pages}{2164} (\bibinfo{year}{1997})\relax
\mciteBstWouldAddEndPuncttrue
\mciteSetBstMidEndSepPunct{\mcitedefaultmidpunct}
{\mcitedefaultendpunct}{\mcitedefaultseppunct}\relax
\EndOfBibitem
\bibitem[{\citenamefont{Vidal}(2004)}]{vidal-2004}
\bibinfo{author}{\bibfnamefont{G.}~\bibnamefont{Vidal}},
  \bibinfo{journal}{Phys. Rev. Lett.} \textbf{\bibinfo{volume}{93}},
  \bibinfo{pages}{040502} (\bibinfo{year}{2004})\relax
\mciteBstWouldAddEndPuncttrue
\mciteSetBstMidEndSepPunct{\mcitedefaultmidpunct}
{\mcitedefaultendpunct}{\mcitedefaultseppunct}\relax
\EndOfBibitem
\bibitem[{\citenamefont{Verstraete et~al.}(2004)\citenamefont{Verstraete,
  Porras, and Cirac}}]{frank-2004a}
\bibinfo{author}{\bibfnamefont{F.}~\bibnamefont{Verstraete}},
  \bibinfo{author}{\bibfnamefont{D.}~\bibnamefont{Porras}}, \bibnamefont{and}
  \bibinfo{author}{\bibfnamefont{J.~I.} \bibnamefont{Cirac}},
  \bibinfo{journal}{Phys. Rev. Lett.} \textbf{\bibinfo{volume}{93}},
  \bibinfo{pages}{227205} (\bibinfo{year}{2004})\relax
\mciteBstWouldAddEndPuncttrue
\mciteSetBstMidEndSepPunct{\mcitedefaultmidpunct}
{\mcitedefaultendpunct}{\mcitedefaultseppunct}\relax
\EndOfBibitem
\bibitem[{\citenamefont{Verstraete et~al.}(2008)\citenamefont{Verstraete, Murg,
  and Cirac}}]{frank-2008}
\bibinfo{author}{\bibfnamefont{F.}~\bibnamefont{Verstraete}},
  \bibinfo{author}{\bibfnamefont{V.}~\bibnamefont{Murg}}, \bibnamefont{and}
  \bibinfo{author}{\bibfnamefont{J.~I.} \bibnamefont{Cirac}},
  \bibinfo{journal}{Advances in Physics} \textbf{\bibinfo{volume}{57}},
  \bibinfo{pages}{143} (\bibinfo{year}{2008})\relax
\mciteBstWouldAddEndPuncttrue
\mciteSetBstMidEndSepPunct{\mcitedefaultmidpunct}
{\mcitedefaultendpunct}{\mcitedefaultseppunct}\relax
\EndOfBibitem
\bibitem[{\citenamefont{Sandvik and Vidal}(2007)}]{sandvik-vidal-2007}
\bibinfo{author}{\bibfnamefont{A.~W.} \bibnamefont{Sandvik}} \bibnamefont{and}
  \bibinfo{author}{\bibfnamefont{G.}~\bibnamefont{Vidal}},
  \bibinfo{journal}{Phys. Rev. Lett.} \textbf{\bibinfo{volume}{99}},
  \bibinfo{pages}{220602} (\bibinfo{year}{2007})\relax
\mciteBstWouldAddEndPuncttrue
\mciteSetBstMidEndSepPunct{\mcitedefaultmidpunct}
{\mcitedefaultendpunct}{\mcitedefaultseppunct}\relax
\EndOfBibitem
\bibitem[{\citenamefont{Pippan et~al.}(2010)\citenamefont{Pippan, White, and
  Evertz}}]{pippan-2010}
\bibinfo{author}{\bibfnamefont{P.}~\bibnamefont{Pippan}},
  \bibinfo{author}{\bibfnamefont{S.~R.} \bibnamefont{White}}, \bibnamefont{and}
  \bibinfo{author}{\bibfnamefont{H.~G.} \bibnamefont{Evertz}},
  \bibinfo{journal}{Phys. Rev. B} \textbf{\bibinfo{volume}{81}},
  \bibinfo{pages}{081103(R)} (\bibinfo{year}{2010})\relax
\mciteBstWouldAddEndPuncttrue
\mciteSetBstMidEndSepPunct{\mcitedefaultmidpunct}
{\mcitedefaultendpunct}{\mcitedefaultseppunct}\relax
\EndOfBibitem
\bibitem[{\citenamefont{Cardy}(1996)}]{book-cardy-1996}
\bibinfo{author}{\bibfnamefont{J.}~\bibnamefont{Cardy}},
  \emph{\bibinfo{title}{Scaling and renormalization in statistical physics}}
  (\bibinfo{publisher}{Cambridge University Press}, \bibinfo{year}{1996})\relax
\mciteBstWouldAddEndPuncttrue
\mciteSetBstMidEndSepPunct{\mcitedefaultmidpunct}
{\mcitedefaultendpunct}{\mcitedefaultseppunct}\relax
\EndOfBibitem
\bibitem[{\citenamefont{Nishino and Okunishi}(1995)}]{nishino-okunishi-1995}
\bibinfo{author}{\bibfnamefont{T.}~\bibnamefont{Nishino}} \bibnamefont{and}
  \bibinfo{author}{\bibfnamefont{K.}~\bibnamefont{Okunishi}},
  \bibinfo{journal}{Journal of the Physical Society of Japan}
  \textbf{\bibinfo{volume}{64}}, \bibinfo{pages}{4084}
  (\bibinfo{year}{1995})\relax
\mciteBstWouldAddEndPuncttrue
\mciteSetBstMidEndSepPunct{\mcitedefaultmidpunct}
{\mcitedefaultendpunct}{\mcitedefaultseppunct}\relax
\EndOfBibitem
\bibitem[{\citenamefont{Hieida et~al.}(1997)\citenamefont{Hieida, Okunishi, and
  Akutsu}}]{hieida-okunishi-akutsu-1997}
\bibinfo{author}{\bibfnamefont{Y.}~\bibnamefont{Hieida}},
  \bibinfo{author}{\bibfnamefont{K.}~\bibnamefont{Okunishi}}, \bibnamefont{and}
  \bibinfo{author}{\bibfnamefont{Y.}~\bibnamefont{Akutsu}},
  \bibinfo{journal}{Physics Letters A} \textbf{\bibinfo{volume}{233}},
  \bibinfo{pages}{464 } (\bibinfo{year}{1997}), ISSN
  \bibinfo{issn}{0375-9601}\relax
\mciteBstWouldAddEndPuncttrue
\mciteSetBstMidEndSepPunct{\mcitedefaultmidpunct}
{\mcitedefaultendpunct}{\mcitedefaultseppunct}\relax
\EndOfBibitem
\bibitem[{\citenamefont{Okunishi et~al.}(1999)\citenamefont{Okunishi, Hieida,
  and Akutsu}}]{okunishi-hieida-akutsu-1999}
\bibinfo{author}{\bibfnamefont{K.}~\bibnamefont{Okunishi}},
  \bibinfo{author}{\bibfnamefont{Y.}~\bibnamefont{Hieida}}, \bibnamefont{and}
  \bibinfo{author}{\bibfnamefont{Y.}~\bibnamefont{Akutsu}},
  \bibinfo{journal}{Phys. Rev. E} \textbf{\bibinfo{volume}{59}},
  \bibinfo{pages}{R6227} (\bibinfo{year}{1999})\relax
\mciteBstWouldAddEndPuncttrue
\mciteSetBstMidEndSepPunct{\mcitedefaultmidpunct}
{\mcitedefaultendpunct}{\mcitedefaultseppunct}\relax
\EndOfBibitem
\bibitem[{\citenamefont{Ueda et~al.}(2006)\citenamefont{Ueda, Nishino,
  Okunishi, Hieida, Derian, and
  Gendiar}}]{ueda-nishino-okunishi-hieida-derian-gendiar-2006}
\bibinfo{author}{\bibfnamefont{K.}~\bibnamefont{Ueda}},
  \bibinfo{author}{\bibfnamefont{T.}~\bibnamefont{Nishino}},
  \bibinfo{author}{\bibfnamefont{K.}~\bibnamefont{Okunishi}},
  \bibinfo{author}{\bibfnamefont{Y.}~\bibnamefont{Hieida}},
  \bibinfo{author}{\bibfnamefont{R.}~\bibnamefont{Derian}}, \bibnamefont{and}
  \bibinfo{author}{\bibfnamefont{A.}~\bibnamefont{Gendiar}},
  \bibinfo{journal}{Journal of the Physical Society of Japan}
  \textbf{\bibinfo{volume}{75}}, \bibinfo{pages}{014003}
  (\bibinfo{year}{2006})\relax
\mciteBstWouldAddEndPuncttrue
\mciteSetBstMidEndSepPunct{\mcitedefaultmidpunct}
{\mcitedefaultendpunct}{\mcitedefaultseppunct}\relax
\EndOfBibitem
\bibitem[{\citenamefont{Vidal}(2007)}]{vidal-2007}
\bibinfo{author}{\bibfnamefont{G.}~\bibnamefont{Vidal}},
  \bibinfo{journal}{Phys. Rev. Lett.} \textbf{\bibinfo{volume}{98}},
  \bibinfo{pages}{070201} (\bibinfo{year}{2007})\relax
\mciteBstWouldAddEndPuncttrue
\mciteSetBstMidEndSepPunct{\mcitedefaultmidpunct}
{\mcitedefaultendpunct}{\mcitedefaultseppunct}\relax
\EndOfBibitem
\bibitem[{\citenamefont{McCulloch}(2008)}]{mcculloch-2008}
\bibinfo{author}{\bibfnamefont{I.~P.} \bibnamefont{McCulloch}}
  (\bibinfo{year}{2008}), \eprint{arXiv:0804.2509v1}\relax
\mciteBstWouldAddEndPuncttrue
\mciteSetBstMidEndSepPunct{\mcitedefaultmidpunct}
{\mcitedefaultendpunct}{\mcitedefaultseppunct}\relax
\EndOfBibitem
\bibitem[{\citenamefont{Or\'us and Vidal}(2008)}]{orus-2008}
\bibinfo{author}{\bibfnamefont{R.}~\bibnamefont{Or\'us}} \bibnamefont{and}
  \bibinfo{author}{\bibfnamefont{G.}~\bibnamefont{Vidal}},
  \bibinfo{journal}{Phys. Rev. B} \textbf{\bibinfo{volume}{78}},
  \bibinfo{pages}{155117} (\bibinfo{year}{2008})\relax
\mciteBstWouldAddEndPuncttrue
\mciteSetBstMidEndSepPunct{\mcitedefaultmidpunct}
{\mcitedefaultendpunct}{\mcitedefaultseppunct}\relax
\EndOfBibitem
\bibitem[{\citenamefont{Pirvu et~al.}(2010)\citenamefont{Pirvu, Murg, Cirac,
  and Verstraete}}]{me-2010}
\bibinfo{author}{\bibfnamefont{B.}~\bibnamefont{Pirvu}},
  \bibinfo{author}{\bibfnamefont{V.}~\bibnamefont{Murg}},
  \bibinfo{author}{\bibfnamefont{J.~I.} \bibnamefont{Cirac}}, \bibnamefont{and}
  \bibinfo{author}{\bibfnamefont{F.}~\bibnamefont{Verstraete}},
  \bibinfo{journal}{New J. Phys.} \textbf{\bibinfo{volume}{12}},
  \bibinfo{pages}{025012} (\bibinfo{year}{2010})\relax
\mciteBstWouldAddEndPuncttrue
\mciteSetBstMidEndSepPunct{\mcitedefaultmidpunct}
{\mcitedefaultendpunct}{\mcitedefaultseppunct}\relax
\EndOfBibitem
\bibitem[{\citenamefont{Nishino et~al.}(1996)\citenamefont{Nishino, Okunishi,
  and Kikuchi}}]{nishino-1996}
\bibinfo{author}{\bibfnamefont{T.}~\bibnamefont{Nishino}},
  \bibinfo{author}{\bibfnamefont{K.}~\bibnamefont{Okunishi}}, \bibnamefont{and}
  \bibinfo{author}{\bibfnamefont{M.}~\bibnamefont{Kikuchi}},
  \bibinfo{journal}{Physics Letters A} \textbf{\bibinfo{volume}{213}},
  \bibinfo{pages}{69 } (\bibinfo{year}{1996}), ISSN
  \bibinfo{issn}{0375-9601}\relax
\mciteBstWouldAddEndPuncttrue
\mciteSetBstMidEndSepPunct{\mcitedefaultmidpunct}
{\mcitedefaultendpunct}{\mcitedefaultseppunct}\relax
\EndOfBibitem
\bibitem[{\citenamefont{Andersson et~al.}(1999)\citenamefont{Andersson, Boman,
  and \"Ostlund}}]{andersson-boman-ostlund-1999}
\bibinfo{author}{\bibfnamefont{M.}~\bibnamefont{Andersson}},
  \bibinfo{author}{\bibfnamefont{M.}~\bibnamefont{Boman}}, \bibnamefont{and}
  \bibinfo{author}{\bibfnamefont{S.}~\bibnamefont{\"Ostlund}},
  \bibinfo{journal}{Phys. Rev. B} \textbf{\bibinfo{volume}{59}},
  \bibinfo{pages}{10493} (\bibinfo{year}{1999})\relax
\mciteBstWouldAddEndPuncttrue
\mciteSetBstMidEndSepPunct{\mcitedefaultmidpunct}
{\mcitedefaultendpunct}{\mcitedefaultseppunct}\relax
\EndOfBibitem
\bibitem[{\citenamefont{Tagliacozzo et~al.}(2008)\citenamefont{Tagliacozzo,
  de~Oliveira, Iblisdir, and Latorre}}]{tagliacozzo-2008}
\bibinfo{author}{\bibfnamefont{L.}~\bibnamefont{Tagliacozzo}},
  \bibinfo{author}{\bibfnamefont{T.~R.} \bibnamefont{de~Oliveira}},
  \bibinfo{author}{\bibfnamefont{S.}~\bibnamefont{Iblisdir}}, \bibnamefont{and}
  \bibinfo{author}{\bibfnamefont{J.~I.} \bibnamefont{Latorre}},
  \bibinfo{journal}{Phys. Rev. B} \textbf{\bibinfo{volume}{78}},
  \bibinfo{eid}{024410} (\bibinfo{year}{2008})\relax
\mciteBstWouldAddEndPuncttrue
\mciteSetBstMidEndSepPunct{\mcitedefaultmidpunct}
{\mcitedefaultendpunct}{\mcitedefaultseppunct}\relax
\EndOfBibitem
\bibitem[{\citenamefont{Pollmann et~al.}(2009)\citenamefont{Pollmann, Mukerjee,
  Turner, and Moore}}]{moore-2008}
\bibinfo{author}{\bibfnamefont{F.}~\bibnamefont{Pollmann}},
  \bibinfo{author}{\bibfnamefont{S.}~\bibnamefont{Mukerjee}},
  \bibinfo{author}{\bibfnamefont{A.~M.} \bibnamefont{Turner}},
  \bibnamefont{and} \bibinfo{author}{\bibfnamefont{J.~E.} \bibnamefont{Moore}},
  \bibinfo{journal}{Phys. Rev. Lett.} \textbf{\bibinfo{volume}{102}},
  \bibinfo{pages}{255701} (\bibinfo{year}{2009})\relax
\mciteBstWouldAddEndPuncttrue
\mciteSetBstMidEndSepPunct{\mcitedefaultmidpunct}
{\mcitedefaultendpunct}{\mcitedefaultseppunct}\relax
\EndOfBibitem
\bibitem[{\citenamefont{Pirvu~\emph{et al}.}(2010)}]{inprep-2010}
\bibinfo{author}{\bibfnamefont{B.}~\bibnamefont{Pirvu~\emph{et al}.}},
  \bibinfo{journal}{in preparation}  (\bibinfo{year}{2010})\relax
\mciteBstWouldAddEndPuncttrue
\mciteSetBstMidEndSepPunct{\mcitedefaultmidpunct}
{\mcitedefaultendpunct}{\mcitedefaultseppunct}\relax
\EndOfBibitem
\bibitem[{\citenamefont{Press et~al.}(2007)\citenamefont{Press, Teukolsky,
  Vetterling, and Flannery}}]{numrec-2007}
\bibinfo{author}{\bibfnamefont{W.~H.} \bibnamefont{Press}},
  \bibinfo{author}{\bibfnamefont{S.~A.} \bibnamefont{Teukolsky}},
  \bibinfo{author}{\bibfnamefont{W.~T.} \bibnamefont{Vetterling}},
  \bibnamefont{and} \bibinfo{author}{\bibfnamefont{B.~P.}
  \bibnamefont{Flannery}}, \emph{\bibinfo{title}{Numerical Recipes: The Art of
  Scientific Computing}} (\bibinfo{publisher}{Cambridge University Press},
  \bibinfo{year}{2007})\relax
\mciteBstWouldAddEndPuncttrue
\mciteSetBstMidEndSepPunct{\mcitedefaultmidpunct}
{\mcitedefaultendpunct}{\mcitedefaultseppunct}\relax
\EndOfBibitem
\bibitem[{\citenamefont{Lieb et~al.}(1961)\citenamefont{Lieb, Schultz, and
  Mattis}}]{lsm-1961}
\bibinfo{author}{\bibfnamefont{E.}~\bibnamefont{Lieb}},
  \bibinfo{author}{\bibfnamefont{T.}~\bibnamefont{Schultz}}, \bibnamefont{and}
  \bibinfo{author}{\bibfnamefont{D.}~\bibnamefont{Mattis}},
  \bibinfo{journal}{Annals of Physics} \textbf{\bibinfo{volume}{16}},
  \bibinfo{pages}{407 } (\bibinfo{year}{1961})\relax
\mciteBstWouldAddEndPuncttrue
\mciteSetBstMidEndSepPunct{\mcitedefaultmidpunct}
{\mcitedefaultendpunct}{\mcitedefaultseppunct}\relax
\EndOfBibitem
\bibitem[{\citenamefont{White and Huse}(1993)}]{white-huse-1993}
\bibinfo{author}{\bibfnamefont{S.~R.} \bibnamefont{White}} \bibnamefont{and}
  \bibinfo{author}{\bibfnamefont{D.~A.} \bibnamefont{Huse}},
  \bibinfo{journal}{Phys. Rev. B} \textbf{\bibinfo{volume}{48}},
  \bibinfo{pages}{3844} (\bibinfo{year}{1993})\relax
\mciteBstWouldAddEndPuncttrue
\mciteSetBstMidEndSepPunct{\mcitedefaultmidpunct}
{\mcitedefaultendpunct}{\mcitedefaultseppunct}\relax
\EndOfBibitem
\end{mcitethebibliography}


\end{document}